\documentclass[11pt,a4paper,pdftex]{article}
\pdfoutput=1
\usepackage{jcappub}
\usepackage{bm}

\usepackage{epsfig}
\usepackage{citesort}
\usepackage{graphicx}
\usepackage{amsmath}
\usepackage{amssymb}
\usepackage{amsbsy}
\usepackage{color}
\usepackage{subfigure}
\usepackage{slashed}
\usepackage{afterpage}
\usepackage{psfrag}

\usepackage{comment}

\renewcommand\({\left(}
\renewcommand\){\right)}

\newcommand{\be}{\begin{equation}}
\newcommand{\ee}{\end{equation}}
\newcommand{\bea}{\begin{eqnarray}}
\newcommand{\eea}{\end{eqnarray}}
\newcommand{\bb}{\begin{equation}}
\newcommand{\eqb}{\begin{eqnarray}}
\newcommand{\eqf}{\end{eqnarray}}
\def\nn{\nonumber}

\newcommand{\A}{\mu}

\renewcommand\({\left(}
\renewcommand\){\right)}

\def\bra{\langle}
\def\ket{\rangle}

\newcommand{\muu}{{m_{\gamma'}}}

\newcommand{\exclude}[1]{}

\definecolor{gre}{rgb}{0,0.4,0.3}
\definecolor{durbeer}{rgb}{1,0,0.3}

\definecolor{crap}{rgb}{0.5,0.1,0.7}

\begin{document}
\subheader{\rm\hfill DESY 11-226; MPP-2011-140; CERN-PH-TH/2011-323; IPPP/11/80; DCPT/11/160}

\title{WISPy Cold Dark Matter}

\author[a,b]{Paola~Arias}
\author[c]{Davide~Cadamuro}
\author[a,d]{Mark~Goodsell}
\author[e]{Joerg~Jaeckel} 
\author[c]{Javier~Redondo}
\author[a]{Andreas~Ringwald}

\affiliation[a]{Deutsches Elektronen-Synchrotron, Notkestra\ss e 85, D-22607 Hamburg, Germany}
\affiliation[b]{Facultad de F\'isica, Pontificia Universidad Cat\'olica de Chile, Casilla 306, Santiago 22, Chile}
\affiliation[c]{Max-Planck-Institut f\"ur Physik, 
F\"ohringer Ring 6, D-80805 M\"unchen, Germany}
\affiliation[d]{Department of Physics, CERN Theory Division, CH-1211 Geneva 23, Switzerland}
\affiliation[e]{Institute for Particle Physics Phenomenology, Durham University, Durham DH1 3LE, UK}

\emailAdd{paola.arias@desy.de}
\emailAdd{cadamuro@mppmu.mpg.de}
\emailAdd{mark.goodsell@cern.ch}
\emailAdd{joerg.jaeckel@durham.ac.uk}
\emailAdd{redondo@mppmu.mpg.de}
\emailAdd{andreas.ringwald@desy.de}

\abstract{Very weakly interacting slim particles (WISPs), such as axion-like particles (ALPs) or hidden photons (HPs), may be non-thermally produced via the misalignment mechanism in the early universe and survive as a cold dark matter population until today. 
We find that, both for ALPs and HPs whose dominant interactions with the standard model arise from couplings to photons, a huge region in the parameter spaces spanned by photon coupling and ALP or HP mass can give rise to the observed cold dark matter.
Remarkably, a large region of this parameter space coincides with that predicted in well motivated models of fundamental physics.
A wide range of experimental searches -- exploiting haloscopes (direct dark matter searches exploiting microwave cavities), helioscopes (searches for solar ALPs or HPs),  or light-shining-through-a-wall techniques --  can probe large parts of this parameter space in the foreseeable future. }

\maketitle

\section{\label{intro}Introduction}

The two most relevant features of dark matter (DM) particles are their feeble interactions with standard model particles 
and their cosmological stability. In addition DM is required to be sufficiently cold in order to allow for efficient structure formation.

A realization of all these features are weakly interacting massive particles (WIMPs). Thermally produced in the early Universe their large, of order TeV scale mass ensures that by now they are non-relativistic and therefore sufficiently cold. Their interactions are small due to the large mass of the mediator particles (such as W or Z bosons) which makes the interaction very short ranged. Despite this, their large mass and the correspondingly large phase space is at odds with the required stability on cosmological time-scales. In order to ensure this stability one is therefore forced to introduce symmetries that conserve the number of these DM particles.
However, motivating these symmetries on theoretical grounds is non-trivial. Global symmetries may be broken in quantum gravity whereas local symmetries lead to additional interactions which may cause conflicts with the required weakness of the DM interactions.
Nevertheless good candidates exist. Two of the most famous examples  
are: the lightest supersymmetric particle in supersymmetric models with R-parity (see, e.g.~\cite{Jungman:1995df} for a review), or the lightest Kaluza-Klein modes in models with conserved parity in extra dimensions (see, e.g.~\cite{Hooper:2007qk} for a review), both models very appealing because of their connection with more fundamental theories of space and time, and their discovery potential at the LHC. 

Although it is way too early to make a final judgement it is nevertheless noteworthy that the initial searches at LHC have not given any indication of the existence of WIMPs. Because of this and the mentioned theoretical issues it is interesting and timely to consider alternative ways to realize the essential features of DM.

Sufficient stability of the DM particles can also be achieved by combining the weakness of their interactions with a sufficiently small mass. The latter drastically reduces the phase space (and the number/type of possible decay products) thereby increasing the lifetime. This is the road we want to pursue in this paper: we will concentrate on very weakly interacting slim particles (WISPs) as DM candidates.

Yet, a thermally produced light DM candidate can run foul of the required coldness of DM and can prevent structure formation. 
More precisely, the free-streaming length (the distance a DM particle can travel before getting trapped in a potential well) would increase with decreasing mass, and therefore at some point these DM particles would be inconsistent with the existence of dwarf galaxies, galaxies, clusters, superclusters and so on. This argument can be used for instance to rule out standard neutrinos as DM. 
This reasoning is extremely powerful in light of the increasingly precise cosmological data at our disposal and even subdominant components of thermally produced light DM can be ruled out, the case of eV mass axions being a prime example (see, e.g.~\cite{Hannestad:2010yi}).

However, there are non-thermal means for producing sufficiently cold dark matter made of light particles. 
One of the most interesting is the misalignment mechanism, discussed mostly in the case of the QCD axion~\cite{Preskill:1982cy,Abbott:1982af,Dine:1982ah} or (recently) string axions~\cite{arXiv:0905.4720,arXiv:1004.5138,Marsh:2011gr,Higaki:2011me} and the central topic of this paper. Very recently, this mechanism has also been proposed to produce cold dark matter (CDM) out of hidden photons (HPs)~\cite{arXiv:1105.2812}.

In this paper, we shall revisit the misalignment mechanism for both cases: we treat first axion-like particles (ALPs - which may arise as pseudo-Nambu-Goldstone bosons in field theory and appear generically in all string compactifications) and then hidden photons. 
Our conclusions turn out to be extremely motivating. 
Once produced, a population of very light cold dark matter particles is extremely difficult to reabsorb by the primordial plasma. Therefore, we find that in both cases, ALPs and HPs, 
a huge region in parameter space spanned by their masses and their couplings to standard model particles 
can give rise to the observed dark matter.  
These topics have already been discussed in some detail in~\cite{Masso:2004cv,arXiv:1105.2812}.  
The novelty in this work is that we shall provide new constraints and expose interesting regions of parameter space relevant for direct and indirect searches.

The outline of this paper goes as follows: in Sect.~\ref{essentials} we review the essentials of the misalignment mechanism. 
In Sects.~\ref{sec:ALPs} and~\ref{sec:HPs} we elaborate on two particular cases, axion-like particles and hidden photons, respectively.
We discuss the cosmological constraints, also noting some misconceptions in the results of~\cite{arXiv:1105.2812}. In Sect.~\ref{sec:direct} we discuss the direct detection of ALP and HP CDM in microwave cavity experiments. We conclude in Sect.~\ref{conclusions}.

\section{\label{essentials}Essentials of the misalignment mechanism}

The misalignment mechanism relies on assuming that fields in the early universe have a random initial state 
(such as as one would expect, for example, to arise from quantum fluctuations during inflation) which is fixed by the expansion of the universe; fields with mass $m$ evolve on timescales $t \sim m^{-1}$. After such a timescale, the fields respond by attempting to minimize their potential, and consequently oscillate around the minimum. If there is no significant damping via decays, these oscillations can behave as a cold dark matter fluid since their energy density is diluted by the expansion of the universe as $\rho\propto a^{-3}$, where $a$ is the universe scale factor.

In order to be more quantitative, let us revisit this mechanism for the simple case of a real scalar field with Lagrangian,
\bb 
\mathcal L= \frac{1}{2}\partial_\mu \phi \partial^\mu \phi-\frac{1}{2}m_{\phi}^2\phi^2 +{\cal L}_I ,  
\ee
where ${\cal L}_I$ encodes interactions of the scalar field with itself and the rest of particles in the primordial bath. 
We assume that the universe underwent a period of inflation at a value of the Hubble expansion parameter $H=d \log a/dt$ larger than the scalar mass. 
After inflation the field shall be approximately spatially uniform and the initial state is characterized by a single initial value, $\phi_i$. 
After inflation a period of reheating occurs, followed by a period of radiation dominated expansion. 
The  equation of motion for $\phi$ in the expanding Universe is
\begin{equation}
\label{eq:condensev}
\ddot{\phi}+3H\dot{\phi}+m_{\phi}^{2}\phi=0 .
\end{equation}  
In general, the mass receives thermal corrections from ${\cal L}_I$ which might be crucial, thus $m_{\phi}=m_{\phi}(t)$ should be understood. 

The solution of this equation can be separated into two regimes. In a first epoch, $3H\gg m_{\phi}$, so $\phi$ is an overdamped oscillator and gets frozen, $\dot \phi=0 $. 
At a later time, $t_1$, characterized by $3H(t_1)=m_{\phi}(t_1)\equiv m_1$, the damping becomes undercritical and the field can roll down the potential and start to oscillate. During this epoch, the mass term is the leading scale in the equation and the solution can be found in the WKB approximation,
\be
\label{eq:WKB}
\phi\simeq \phi_1 \(\frac{m_1 a_1^3}{m_{\phi} a^3}\)^{1/2} \cos\(\int_{t_1}^t m_{\phi}\, dt\) ,  
\ee
where $\phi_1\sim \phi_i$ since up to $t_1$ the evolution is frozen. 
Note that in obtaining this solution we have not assumed a particular form for $H$ but just its definition, and so it is valid for the radiation, matter, and vacuum energy dominated phases of the universe and their transitions. 

The solution corresponds to fast oscillations with a slow amplitude decay. 
Let us call this amplitude ${\cal A}(t)=\phi_1(m_1 a_1^3/m_{\phi} a^3)^{1/2}$ and the phase $\alpha(t)=\int^t m_{\phi}(t)dt$. 
The energy density of the scalar field is
\be
\label{scalardensity}
\rho_{\phi} =  \frac{1}{2}\dot \phi^2+\frac{1}{2}m^2_{\phi}\phi^2 = \frac{1}{2}m^2_{\phi}{\cal A}^2 + ...\ ,
\ee
where the dots stand for terms involving derivatives of ${\cal A}$, which by assumption are much smaller than $m_{\phi}$ ($m_{\phi}\gg H$ in this regime). 
Let us also consider the pressure,
\be
p_{\phi}    = \frac{1}{2}\dot \phi^2-\frac{1}{2}m_{\phi}^2\phi^2 =-\frac{1}{2}m_{\phi}^2{\cal A}^2\cos\(2\alpha\)-{\cal A}\dot{\cal A} m_{\phi} \sin\(2\alpha\)+\dot {\cal A}^2\cos^2\(\alpha\).
\ee
At times $t\gg t_{1}$\footnote{When the field just starts to oscillate the averaging employed in the following is not a good approximation and the equation of state is a non-trivial and strongly time dependent function. Depending on when the transition occurs this may have interesting cosmological effects on, e.g., structure formation.} the oscillations in the pressure occur at time scales $1/m_{\phi}$ much much faster than the cosmological evolution.
We can therefore average over these oscillations, giving
\begin{equation}
\langle p_{\phi} \rangle   = \langle\dot {\cal A}^2\cos^2\(\alpha\)\rangle=\frac{1}{2}\dot{\cal A}^2.
\end{equation}
As already mentioned $\dot{\cal A}\ll m_{\phi}{\cal A}$. Thus, at leading order in $\dot{\cal A}/({\cal A}m)$, the equation of state is just
\begin{equation}  
w=\langle p\rangle/\langle \rho\rangle \simeq 0,
\end{equation}
which is exactly that of non-relativistic matter. 

It follows from  \eqref{eq:WKB} that the energy density in a comoving volume, $\rho a^3$, is not conserved if the scalar mass 
changes in time. The quantity 
\begin{equation}
N=\rho a^3/m_{\phi}=\frac{1}{2}m_{1} a^{3}_{1}\phi^{2}_{1},
\end{equation} 
is however constant even in this case, and can be interpreted as a comoving number of non-relativistic quanta of mass $m_{\phi}$.
Here, we only need the conservation of $N$ to compute the energy density today as
\begin{equation}
\label{mos}
\rho_{\phi} (t_0)=m_0 \frac{N}{a_0^3}\simeq
 \frac{1}{2}m_0 m_1 \phi_1^2\left(\frac{a_1}{a_0}\right)^{3} , 
\end{equation}
where quantities with a $0$-subscript are evaluated at present time. 

More physics insight is gained using temperatures instead of times and scale factors. 
First, we use the conservation of comoving entropy $S=s a^3=2\pi g_{*S}(T)T^3 a^3/45$ to write $(a_1/a)^3=g_{*S}(T)T^3/g_{*S}(T_1)T_1^3$. 
Then we use the expression for the Hubble constant in the radiation dominated era $H=1.66 \sqrt{g_*(T)}T^2/m_{\rm Pl}$ and the definition of $T_1$, $3H(T_1)=m_1$ to express $T_1$ in terms of $m_1$ and the Planck mass $m_{\rm Pl}=1.22\times10^{19}$~GeV. The functions $g_*$ and $g_{*S}$ are the effective numbers of energy and entropy degrees of freedom defined in~\cite{Kolb}. 
The dark matter density today, (\ref{mos}), can then be expressed as 
\be 
\label{eq:CCDM}
\rho_{\phi,0}\simeq 0.17\, \frac{{\rm keV}}{{\rm cm}^3}\times \sqrt{\frac{m_0}{{\rm eV}}} \sqrt{\frac{m_0}{m_1}}\(\frac{\phi_1}{10^{11}\, {\rm GeV}}\)^2 {\cal F}(T_1) ,
\ee
where ${\cal F}(T_1)\equiv (g_{*}(T_1)/3.36)^\frac{3}{4}(g_{*S}(T_1)/3.91)^{-1}$ is a smooth function ranging from $1$ to $\sim 0.3$ in the interval 
$T_1\in (T_0, 200 {\rm GeV})$.
The abundance is most sensitive to the initial amplitude of the oscillations, $\propto \phi_1^2$, and to a lesser degree to today's mass $m_0$. 
The factor $\propto 1/\sqrt{m_1}$ reflects the damping of the oscillations in the expanding universe:  the later the oscillations start, i.e.~the smaller $T_1$ and therefore $H_1$ and $m_1$, the less damped they are for a given $m_0$. 

If we compare the above estimate with the DM density measured by WMAP and other large scale structure probes~\cite{arXiv:1001.4538},
\be
\label{eq:observedDM}
\rho_{\rm CDM} = {1.17(6)} \frac{{\rm keV}}{{\rm cm}^3} , 
\ee
it is clear that we need very large values of $\phi_1$ to account for all the dark matter. 
However, a relatively small $\phi_1$ could be compensated by a small $m_1\ll m_{0}$.  

If we want the condensate to mimic the behaviour of standard cold dark matter we should ensure 
that, at latest at  matter-radiation equality, at a temperature $T_{\rm eq}\sim 1.3$~eV, 
the mass attains its current value $m_0$ and therefore the DM density starts to scale truly as $1/a^3$. 
In particular, at this point the field should already have started to oscillate.
This corresponds to a lower limit\footnote{{See also~\cite{Das:2006ht}.}} on $m_1$, $m_1>3 H(T_{\rm eq})=1.8\times 10^{-27}$ eV, which implies an upper 
bound on  $\rho_{\phi,0}$,
\be
\label{eq:ALPDMbound}
\rho_{\phi,0}<  {1.17}\, \frac{{\rm keV}}{{\rm cm}^3}\times \frac{m_0}{{\rm eV}}\(\frac{\phi_1}{{53}\, {\rm TeV}}\)^2 .
\ee
In other words if we want these particles to be the DM, we need that $(m_0/{\rm eV})(\phi_1/{53}\,{\rm TeV})^2>1$,  giving us a constraint on the required initial field value as a function of the mass today.

To conclude this section let us note that dark matter generated by the misalignment mechanism may have interesting properties beyond those of cold dark matter. At the time of their production, particles from the misalignment mechanism are semi-relativistic. Their momenta are of the order of the Hubble constant $p\sim H_{1}\ll T_{1}$; accordingly we have (outside of gravitational wells) a velocity distribution with a very narrow width of roughly,
\begin{equation}
\delta v(t)\sim \frac{H_{1}}{m_{1}}\left(\frac{a_{1}}{a_{0}}\right)\ll 1.
\end{equation}
Combined with the high number density of particles, $n_{\phi,0}=N/a_0^3=\rho_{\rm CDM}/m_0$, this narrow distribution typically leads to very high occupation numbers for each quantum state,
\begin{equation}
{\cal N}_{\rm occupation}\sim \frac{(2\pi)^3}{4\pi/3}\frac{n_{\phi,0}}{m^{3}_{0}\delta v^3}
\sim 10^{42}\left(\frac{m_{1}}{m_{0}}\right)^{3/2}\left(\frac{\rm eV}{m_{0}}\right)^{5/2},
\end{equation}
where we have used {$a_0/a_1\sim T_1/T_0\sim \sqrt{m_1 m_{\rm Pl}}/T_0$}. 
If the interactions are strong enough to achieve thermalisation, as argued in Refs.~\cite{Sikivie:2009qn,Erken:2011dz} for the case of axions, this high occupation number can lead to the formation of a Bose-Einstein condensate. This could lead to interesting properties which may also lead to interesting signatures in cosmological observations~\cite{Sikivie:2009qn,Erken:2011dz,Sikivie:2010yn,Erken:2011xj,Kain:2011pd}.  Although we will not investigate this intriguing possibility here, we note that these features could also be realized for the more general light DM particles discussed in this paper.

In the following we discuss two particularly interesting possibilities for DM from the misalignment mechanism, axion-like particles and 
hidden photons. 

\section{Axion-like particles}\label{sec:ALPs}

In this section we will focus on a specific type of WISP, namely axion-like particles (ALPs). By this we shall mean particles with only derivative couplings to matter, and in particular an interaction with photons given by
\bb
\mathcal{L} \supset - \frac{1}{4} g \phi F_{\mu \nu} \tilde{F}^{\mu\nu}, 
\label{ALPphotoncoupling}
\ee
where $\phi$ is the ALP and $g$ is a dimensionful coupling. The chief examples of ALPs are pseudo Nambu-Goldstone bosons (pNGBs) and string ``axions'' which can be treated together. 
For concreteness we will focus in this paper on particles of these types.

The cosmology of the ALP condensate depends on the type of interaction generating its mass and in particular how this mass changes through the evolution of the universe.
In the following we will go through a variety of possibilities for the DM formation as well as its cosmological viability for different scenarios. The regions which allow for viable ALP DM are then assembled in Fig.~\ref{fig:ALPDM} in the $m_\phi-g$ plane.

\subsection{Axion-like particles from pNGBs and string theory}

When a continuous global symmetry is spontaneous broken, massless particles appear in the low energy theory: Nambu-Goldstone bosons (NGBs). 
They appear in the Lagrangian as phases of the high energy degrees of freedom. 
Since phases are dimensionless the canonically normalised theory at low energies always involves the combination $\phi/f_\phi$, where $\phi$ is the NGB field and $f_\phi$ is a scale close to the spontaneous symmetry breaking (SSB) scale. 
The range for $\phi/f_\phi$ is $(-\pi,\pi)$ and therefore the natural values for $\phi_1$ are $\sim f_\phi$. String axions on the other hand appear in all compactifications. They share  these properties (having a shift symmetry and being periodic) but with the natural size of $f_\phi$ being the string scale (in type II compactifications this can be somewhat modified by a factor of the typical length scale of the compactification). 

Indeed, all of the global symmetries in the standard model are broken\footnote{Assuming that neutrinos are Majorana fermions, otherwise $B-L$ is an exception.}. Furthermore the black hole no hair theorem and what we know about quantum gravity tell us that this should ultimately also occur to any additional global symmetries. Hence we should have pseudo Nambu-Goldstone Bosons (pNGBs) instead of NGBs. They then have a mass, and can be a dark matter candidate. 

There are many possibilities for breaking the shift symmetry, explicitly or spontaneously, perturbatively or non-perturbatively; 
for stringy axions, the shift symmetry is exact to all orders in perturbation theory and is only broken non-perturbatively, for instance from a non-abelian anomaly, gaugino condensation or stringy instantons. The ALP potential can typically be parametrized as
\begin{align}
\label{ALPpotential}
V(\phi) =& m^2_\phi f^2_\phi \left( 1- \cos \frac{\phi}{f_\phi} \right) .
\end{align}
The mass of the ALP is in general unrelated to the QCD axion mass and in particular will be independent of the temperature unless generated by a sector that is thermalised. 
The ALP will satisfy the equation of motion~\eqref{eq:condensev} as long as $\phi/f_\phi$ is small, i.e.~few oscillations after $t_1$. The inaccuracy of the quadratic approximation can be 
cured by an additional correction factor to \eqref{eq:CCDM}. This is normally an ${\cal O}(1)$ factor except if we fine tune the initial condition to $\phi=\pi f_\phi$.

The dimensionful coupling parameter $g$ in (\ref{ALPphotoncoupling}) can be parametrised as 
\bb
\label{genericgvsf}
g \equiv \frac{\alpha}{2\pi}\frac{1}{f_\phi} {\cal N}.  
\ee
In the simplest case $\mathcal{N}$ is an integer, but this is not true in general when the ALP mixes, either kinetically or via symmetry-breaking effects with other ALPs or with  pseudoscalar mesons. 

We can then represent the allowed regions of ALP dark matter in the $m_\phi$--$g$ plane by using 
\be
\phi_1 =\theta_1 \frac{\alpha {\cal N}}{ 2\pi g}
\ee
with $\theta_1=|\phi_1|/f_\phi$, the initial misalignment angle whose range is restricted to values between $-\pi$ and $\pi$.
The model dependent factor ${\cal N}$ will from now on be taken to be unity, but the reader should keep in mind 
that in principle it can be very different in particular constructions. 

{While we shall adopt a phenomenological approach, showing the allowed region in the $m_\phi-g$ plane, as mentioned above we are motivated by both field theoretical and stringy axion models, and so we may ask what the preferred region of the parameter space is. In both cases, the ALP could be related to the generation of right-handed neutrino masses, and hence have a decay constant at an intermediate scale, i.e. within the cosmological axion window; alternatively it could be associated with a GUT, and have a decay constant at the corresponding scale. For string axions, the \emph{coupling to photons} will be related via a loop factor to either the string scale or the Planck scale (or it could be even weaker - so an ALP with a large coupling would restrict the string scale to be low). For string and field theoretical models the most interesting values are therefore $g \sim (10^{-11},10^{-15}) $ GeV$^{-1}$, $\sim 10^{-19}$ GeV$^{-1}$ and $\sim 10^{-21}$ GeV$^{-1}$ corresponding to intermediate, GUT or Planck scales. 
On the other hand, we have no preconceptions regarding preferred ALP masses; as mentioned above the mass $m_{\phi}$ is expected to be generated by non-perturbative effects, and thus suppressed exponentially relative to some fundamental scale (in IIB string theory it is roughly the gravitino mass which multiplies the exponential factor) so for example with many stringy axions (an ``axiverse" \cite{arXiv:0905.4720,arXiv:1004.5138,Marsh:2011gr,Higaki:2011me}) we could easily populate the allowed parameter region. In the following we will therefore treat both $g$ and $m_{\phi}$ as theoretically unconstrained parameters. }

\subsection{ALP dark matter from the misalignment mechanism}

The value of $\phi_1$ that determines the DM abundance depends on the behaviour of the ALP field during inflation. For a pseudo-Nambu-Goldstone boson, the spontaneous symmetry breaking (SSB) could take place before or after inflation: the pNGB effectively exists only after SSB and it is during the associated phase transition that its initial values are set. For a string axion, provided the inflationary scale is below the string scale (or, equivalently, the decay constant) we should have control over the field theory, and so it will behave like a pNGB with symmetry broken before inflation. 
Assuming its mass to be much smaller than the Hubble scale at the time of SSB, $H_{\rm SSB}$, the ALP field will take random values in different causally disconnected regions of the universe. 
The initial size of these domains cannot be larger than 
\begin{equation}
L_{i, {\rm dom}}\sim \frac{1}{H_{\rm SSB}} \sim \frac{m_{\rm Pl}}{f_\phi^2\sqrt{g_*(f_\phi)}}.
\end{equation}   

For a string axion, or for a pNGB whose associated SSB happens before inflation, the initial domains are stretched over a size larger than the current size of the universe. Consequently the initial field value is the same for every point within our current horizon. The current DM density then depends on this initial field value, leaving an additional parameter to tune the DM density.

On the other hand, if the SSB happens after inflation, the DM density has inhomogeneities of order ${\cal O}(1)$ at length scales $\lesssim L_{i, {\rm dom}}$. Non-linear effects, due to the attractive self-interaction caused by higher order terms in the expansion of the potential~\eqref{ALPpotential}, drive the overabundances to form peculiar DM clumps that are called miniclusters~\cite{267521,hep-ph/9303313,arXiv:astro-ph/9311037,astro-ph/0607341}. 
These miniclusters act like seeds enhancing the successive gravitational clumping that leads to structure formation. 
The minicluster mass is set by the dark matter mass inside the Hubble horizon $d_H=H^{-1}$ when the self-interaction freezes-out, i.e.~$M_{\rm mc}\sim \rho_{\phi}(T_{\lambda})d_H(T_{\lambda})^3$ for the freeze-out temperature $T_{\lambda}$. 
Long-range interactions will be exponentially suppressed at distances longer than $1/m_{\phi}$ so we can expect $T_{\lambda}$ to be of the order  $T_1$, with at most a logarithmic dependence on other parameters. 
This is indeed the case for QCD axions, for which the miniclustering is quenched soon after the QCD phase transition that turns on the potential~\eqref{ALPpotential}~\cite{Sikivie:2009qn} giving $M_{\rm mc}\sim 10^{-12}M_{\odot}$, where $M_{\odot}=1.116\times 10^{57}$ GeV is the solar mass, and a radius $R_{\rm mc}\sim10^{11}$~cm~\cite{Kolb:1995bu}. 
In the case of ALPs, $M_{\rm mc}$ can be larger if  the mass is lighter. 
The authors of~\cite{astro-ph/0607341} pointed out that the present data on the CDM power spectrum constrain $M_{\rm mc}\lesssim 4 \times 10^3 M_{\odot} $ which translates into a lower bound 
in temperature $T_{\lambda}>2\times 10^{-5} {\rm \ GeV}$ and in the ALP mass $m_{\phi}>H(T=2\times 10^{-5} {\rm \ GeV})\sim 10^{-20}$ eV.

If some of these miniclusters survive the tidal disruption during structure formation they should be observable in forthcoming lensing experiments~\cite{astro-ph/0607341,Kolb:1995bu}. 
In any case, at larger scales, the DM density averages to a constant value corresponding to 
$\langle \phi_1^2\rangle \sim \pi^2 f^2_\phi /3 $ bearing the mentioned ${\cal O}(1)$ correction due to the non-harmonic behaviour of large initial phases. 

During the spontaneous symmetry breaking of a global symmetry topological defects such as cosmic strings~\cite{Vilenkin:1982ks} and domain walls can be formed. 
Strings have a thickness $\delta \sim 1/f_\phi$  and typical sizes of the order of the horizon, $L \sim t$. 
As strings enter into the horizon they can rapidly reconnect, form loops and decay into pNGBs.  
If the SSB happens after inflation, we have to consider also their contribution to the energy budget of the universe.
Axions resulting from string decay are known to contribute significantly to their cold DM density, but the exact amount is subject to a long-standing controversy~\cite{Kolb}. 
The debate is focused around the axion emission spectrum. 
Some authors argue that the decay proceeds very fast, with an emission spectrum $1/k$ with high and low energy cutoff of order respectively $1/\delta$ and $1/L$.
In this case the contribution to $n_\phi$ is similar to that from the misalignment mechanism~\cite{Harari:1987ht,Hagmann:1990mj}.
Others put forward that the string decays happen after many oscillations, with a radiation spectrum peaked around $2\pi/L$, which enhances the contribution to cold DM by a multiplicative factor of $\log (L/\delta)\sim \log (f_a/m_a)\sim {\cal O}(60)$~\cite{Davis:1985pt,Davis:1986xc,Vilenkin:1986ku,Davis:1989nj,Dabholkar:1989ju,Battye:1993jv,Battye:1994au}.   
Once the axion potential builds up at the QCD phase transition, domain walls build up. If the axion potential has only one minimum these domain walls can still efficiently decay into axions, leading to a third axion population which is thought {not to be significant in the fast decay scenario, while a recent analysis has shown that it can be the main contribution to the axion energy density in the many oscillations one~\cite{Hiramatsu:2012gg}.  }
If different exactly degenerate vacua exist the domain walls are persistent and can very easily run in conflict with observations. Therefore one assumes a small explicit breaking of the Peccei-Quinn symmetry, which breaks the degeneracy and allows domain walls to decay. It is possible although somehow fine-tuned to do so without compromising the solution to the strong-CP problem. For a recent review on axion cosmology see e.g.~\cite{Sikivie:2006ni}.  

We expect the same type of behavior for ALPs with similar characteristics than axions, i.e. ALPs whose mass is generated at a late phase transition due to a hidden sector which becomes strongly interacting. 
In this case we should keep in mind the controversy of the string decay spectrum and assume an uncertainty of order $\log(f_\phi/m_\phi)$ in the DM abundance. 
The domain wall problem can in principle be solved by strong enough explicit breaking, and their contribution to the DM appears to be subdominant as well. 
In models where the cosmological evolution of the ALP mass is different, the situation can differ from the above. 
These models have to be studied case by case, which is beyond the scope of this paper.  

\begin{figure}[t]
\centering
\includegraphics[scale=0.8]{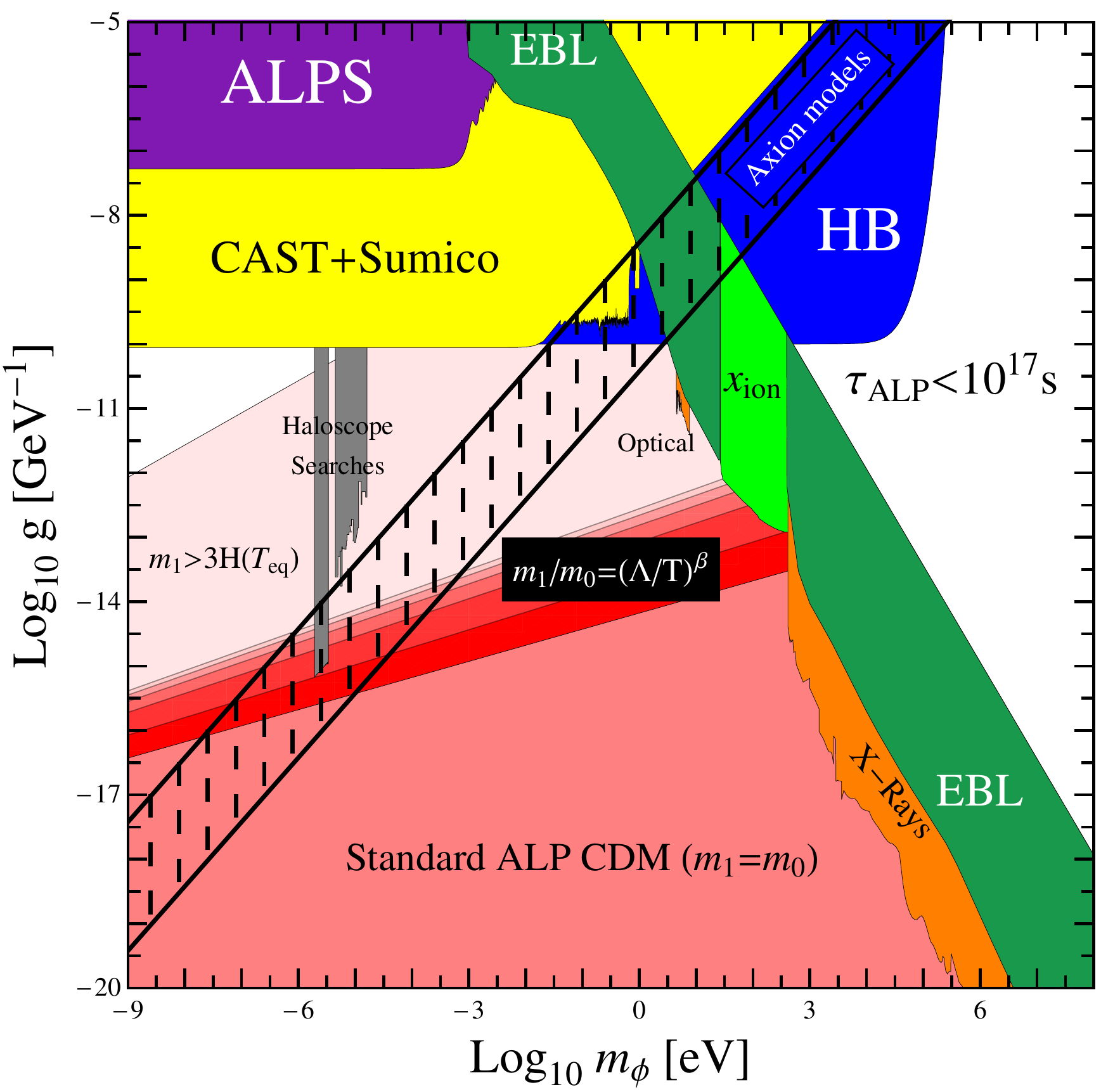}
\caption{Parameter space for axions (shaded band labelled ``Axion models'') and axion-like particles.  The regions where they could form DM are displayed in different shades of red (for details see text). 
The lines representing DM regions are uncertain through a model-dependent multiplicative factor, ${\cal N}$, which we have set equal to 1 here. The DM regions move towards larger couplings $g$, proportional to this factor.
The exclusion regions labelled ``ALPS'', ``CAST+Sumico'' and ``HB'' arise from experiments and astrophysical observations that do not require ALP dark matter (for a review, see~\cite{Jaeckel:2010ni}). The remaining constraints are based on ALPs being DM and are described in the text.
} 
\label{fig:ALPDM}
\end{figure}

\subsection{Sufficient production of dark matter}

As we have seen in Sect.~\ref{essentials} a general constraint arises from the fact that we get a sufficient amount of DM but at the same time 
the mass at matter radiation equality has to be greater than the Hubble constant. This is the bound Eq.~\eqref{eq:ALPDMbound}.
For pNGB ALPs we however also have that the field value itself cannot be larger than $\pi f_{\phi}$ which itself is connected to the 
coupling to photons. Combining these two restrictions gives us a way to constrain the viable regions. 
The light red region in Fig.~\ref{fig:ALPDM} labelled ``$m_1>3H(T_{\rm eq})$'' corresponds to this general bound 
with $\phi_1\leq f_\phi$ and $\cal N$=1.

Any ALP model should satisfy this bound for its zero mode to behave as DM before matter-radiation equality. Realistic models attempting to saturate this bound will have problems either fitting the cosmic microwave background (CMB) data or with the WKB approximation we have used.  
In this sense this bound is very conservative. 
Importantly, for $\mathcal N \sim 1$, it seems to exclude the possibility of providing DM from the type of ultralight ALP field that has been invoked to explain the puzzlingly small opacity of the universe 
for $\sim$ TeV gamma rays (see Ref.~\cite{Horns:2012fx} and references therein) in terms of photon $\leftrightarrow$
ALP conversions in astrophysical magnetic fields, 
requiring\footnote{The required coupling is determined by the extragalactic background light and the size of the astrophysical magnetics fields, therefore plagued by  
sizeable uncertainties.}  $g \sim 10^{-11} {\rm GeV}^{-1}$ and $ m_\phi \lesssim 1$~neV, 
see~\cite{arXiv:0707.4312,arXiv:0712.2825,arXiv:0905.3270,DeAngelis:2011id}. To allow an ALP to explain these observations and simultaneously to be dark matter requires $\mathcal N \gtrsim 10$ which is still conceivable, or a fine tuning of $\theta_1$.
  
Let us now turn to the stronger constraints that can be obtained for specific \linebreak time/temperature dependencies of the ALP mass.
The simplest realization of an ALP model has $m_\phi$ constant throughout the universe expansion. 
In this case, we can infer the DM yield from Eq.~\eqref{eq:CCDM} using $m_1=m_0$. 
The region of ALP DM in this case is depicted in pink in Fig.~\ref{fig:ALPDM} and labelled 
``Standard ALP CDM ($m_1=m_0$)''.  We have assumed ${\cal N}=1$ and used the a-priori unknown value of  $\theta_1$ to tune the right DM abundance. The upper bound on $g$ reflects the fact that $\theta_1$ cannot be larger than $\pi$, and thus assumes $\theta_1\sim \pi$. 
Moving to lower values of $g$ requires inflation happening after SSB in order to have a homogeneous small value of $\theta_1$ which is increasingly fine-tuned to zero. In this sense the values closest to the boundary, corresponding to the largest values of the photon coupling, can be considered the most natural values.

Quantum fluctuations generated in the ALP field during inflation will produce unavoidable inhomogeneities in $\theta_1$ of order $\delta \theta_1 \sim H_I/(2\pi f_\phi)$. This precludes fine-tuning of the universe average of $\phi_1$ below $H_I/2\pi$, and sets a minimum DM abundance for a specified value of $H_I$. 

Since the ALP is effectively massless during inflation, these inhomogeneities correspond to isocurvature perturbations of the gravitational potential; this has been discussed extensively in the literature in the context of axions (see e.g. \cite{arXiv:0904.0647}) and for many string axions, see \cite{arXiv:1004.5138}. The WMAP7 observations of the primordial density fluctuations set very stringent constraints on isocurvature perturbations from which one can obtain an upper bound on $H_I$ assuming a given ALP DM defined essentially by $f_\phi$. Assuming that the decay constant $f_\phi$ does not change during inflation (it certainly should not for string axions, for example) the constraint is \cite{arXiv:1001.4538}
\begin{align}
\alpha \equiv \frac{\bra |S^2| \ket}{\bra |S^2| \ket + \bra |R^2| \ket}< 0.077,
\end{align}
where $\bra |S^2| \ket $ is the isocurvature power spectrum, and $\bra |R^2| \ket$ the adiabatic one (generated by the inflation or other fields). We can approximate $\bra |S^2| \ket \approx \frac{H_I^2}{\pi^2 \phi_1^2}$ where $H_I$ is the Hubble constant during inflation. At the pivot scale $k_0 = 0.002\ \mathrm{MPc}^{-1}$ WMAP finds $\bra |R^2| \ket = 2.42 \times 10^{-9}$ so we have a bound
\begin{align}
H_I
<& {4.5} \times 10^{-5} \phi_1 .
\end{align}
In principle we can constrain the mass of the ALP at the time when it starts oscillating by requiring $3H_I > 3H (T_1) = m_1$, 
\begin{align}
\phi_1 >& {7\times 10^3} m_1, \nn\\
1 
>& {2 \times 10^{-18}} \mathcal{F}^2   \times \left( \frac{{1.17}\ \mathrm{keV cm}^{-3}}{\rho} \right)^2 \left( \frac{m_0}{\mathrm{eV}} \right)^2 \left(\frac{\phi_1}{10^{11}\ \mathrm{GeV}}\right)^3,
\end{align}
{where in the last equation we used (\ref{eq:CCDM}) to get an expression for $m_1$. Therefore, }
at the boundary of sufficient dark matter production when $\phi_1 = \pi f_\phi$, and taking $g_\phi = \alpha/{2\pi f_\phi}$ this translates into
\begin{align}
1>& {0.08}\, \mathcal{F}^2   \times \left( \frac{{1.17}\ \mathrm{keV cm}^{-3}}{\rho} \right)^2 \left( \frac{m_0}{\mathrm{eV}} \right)^2 
\left(\frac{{10^{-19}}\ \mathrm{GeV}^{-1}}{g}\right)^3.
\end{align}
Clearly this is a rather weak constraint, and we are therefore allowed many orders of magnitude between the inverse of the inflationary Hubble constant and the time when the ALP oscillations begin.  Conversely, the reheating temperature $T_{\rm RH}$ is bounded by $T_{\rm RH}^2 \lesssim H_I M_P$; with additional assumptions this can constrain $\phi_1$: for example, requiring leptogenesis with $T_{\rm RH} \sim 10^{9}$ GeV would bound $\phi_1 \gtrsim 10^3$ GeV.
 
We do not have strong arguments against values of {$T_{\rm RH}$} larger than a few MeV~\cite{astro-ph/0403291}, only appealing theoretical prejudices at most, so we shall leave this question aside. 
In any case, the most interesting region from observational purposes corresponds to the largest values of $g$, where the initial angle is not fine-tuned.

This simple ALP model predicts quite weakly coupled ALP CDM. 
Models in which $m_1\ll m_0$ can provide larger DM abundances with smaller values of $f_\phi$ and therefore stronger interactions. 

An interesting example of this case arises if the ALP acquires a mass from coupling to some (hidden) non-abelian group. For a pNGB, this would mean that the associated global symmetry is anomalous under some the non-abelian group, just as the $\eta'$ or the hypothetical axion acquire their mass via QCD instantons.  For our ALP we need in principle another unbroken SU(N) group, which condenses at a scale $\Lambda$.  Then we can parametrize the ALP mass as
\begin{equation}
\label{eq:ALPmassVar}
m_\phi\simeq 
\begin{cases}
\frac{\Lambda'^2}{f_\phi}\equiv m_0&\hbox{for $T\ll \Lambda$},\\
\\
m_0\(\frac{\Lambda''}{T}\)^\beta&\hbox{for $T\gg \Lambda$}.
\end{cases}\ 
\end{equation}  
Here $T$ is the temperature of the new sector. Naively it makes sense to assume $\Lambda'\sim \Lambda''\sim \Lambda$.  
At temperatures larger than $\Lambda$, electric-screening damps long range correlations in the plasma and thus the instantonic configurations, resulting in a decrease of the ALP mass.
In specific models the exponent $\beta$ can be obtained for instance from instanton calculations but here we will treat it as a free parameter.
Assuming the onset of ALP coherent oscillations happens in the mass suppression regime, it is easy to obtain an expression for $m_0/m_1$ which is the expected enhancement in the DM abundance. 
We find
\be
\label{eq:enha1}
\sqrt{\frac{m_0}{m_1}}=\(\frac{\sqrt{m_0 m_{\rm Pl}}}{\Lambda''}\)^\frac{\beta}{\beta+2}\(3\times 1.66\sqrt{g_{*1}}\)^\frac{-\beta}{2\beta+4}
\ee
and the factor that controls the enhancement is
\be
\label{eq:enha2}
\frac{\sqrt{m_0 m_{\rm Pl}}}{\Lambda''}\sim \frac{\Lambda'}{\Lambda''} \sqrt{\frac{m_{\rm Pl}}{f_{\phi}}} .
\ee

Unfortunately these models can provide only a moderate enhancement of the DM density with respect to the constant $m_\phi$ case. The gained regions for the ALP DM case for values of $\beta=1,3,5,7,9$ can be seen in Fig.~\ref{fig:ALPDM} from bottom to top (the lowermost region $m_1=m_0$ corresponds, of course, to $\beta=0$, i.e. to the previously considered case). 
Actually, even considering unrealistic huge values of $\beta$ does not help much, as can be seen from the asymptotic approach of the highest $\beta$ cases. 
This is reflected by the finite limit of Eq.~\eqref{eq:enha1} when $\beta\to \infty$, but it follows from its definition, Eq.~\eqref{eq:ALPmassVar}. 
In the $\beta\to \infty$ limit $m_\phi$ is a step function of temperature $m_\phi=\Theta(T-\Lambda)$, and the relation $m_0=\Lambda^2/f_\phi$ determines $\Lambda$ from $m_0$ and $f_\phi$ (or  $g$) so each point in the $m_\phi-g$ parameter space has an implicit maximum DM abundance, independent of $\beta$.   

Surprisingly, it appears that the crucial assumption that leads to these conclusions is that 
$\Lambda'\sim \Lambda$ because it does not allow to consider arbitrary small values for $\Lambda$ for a given mass. Therefore, models in which $\Lambda'\gg \Lambda$ imply generically higher DM abundance and therefore require smaller initial amplitudes $\phi_1$ and consequently stronger interactions more prone to discovery.
Unfortunately, at the moment we cannot provide for a fully motivated example. 

Finally let us note that the hidden sector responsible for the thermal mass of the ALP can have implications for cosmology if it survives until SM temperatures below the MeV range. For instance it can behave as dark radiation or dark matter during BBN. These constraints have to be studied case by case but can be easily avoided if the temperature of the hidden sector is smaller than the SM bath. 

\subsection{Survival of the condensate}

The ALP CDM scenario can be tested via its coupling to photons.   However, a blessing for detection purposes might also be the model's curse. 
Firstly, the two photon coupling endows ALPs with the decay channel $\phi\to \gamma\gamma$. 
The corresponding lifetime in vacuum\footnote{In the presence of a photon thermal bath the decay is stimulated by a factor $1/(1-e^{-m_\phi/2T})^2$.} is 
\bb
\tau_\phi\equiv \frac{1}{\Gamma_{\gamma\gamma} }= \frac{64\pi}{g^2 m_\phi^3} 
\approx
{1.3\times 10^{25}\, {\rm s}}  \left( \frac{g}{10^{-10}\ \mathrm{GeV}^{-1}} \right)^{-2} \left(\frac{m_\phi}{\mathrm{eV}} \right)^{-3}.   
\ee
ALPs with a  lifetime shorter than the age of the universe, $\sim 13.7 $ Gyr, cannot account for the DM observed in galaxies and have to be discarded. In the $m_\phi-g$ plane this corresponds to the region in the up-right corner
of Fig.~\ref{fig:ALPDM}, labelled ``$\tau<10^{17}{\rm s}$'' and excluded from further discussion in this paper.
Even if ALPs have much longer lifetimes, the few decay photons can still be significantly above the measured photon backgrounds. We explore this possibility later on. 

Secondly, ALPs from the condensate can be absorbed by a thermal photon which is either on-shell ($\gamma \phi \to \gamma^*$) or off-shell ($\gamma^* \phi \to \gamma$). 
Off-shell photons $\gamma^*$ are understood to be absorbed or emitted by another participating particle. 
The inverse-Primakoff process is a notable example of the latter, with the extra particle being a charged particle from the plasma, for instance an electron.  

The thermalization rate of ALPs due to the Primakoff process has been treated in the literature and found to be~\cite{Braaten:1991dd,Bolz:2000fu}
\be
\label{eq:Prima}
\Gamma_{\phi q^\pm}=\frac{\alpha\, g^2}{12}T^3\(\log\(\frac{T^2}{m_\gamma^2}\)+0.8194\) . 
\ee
This is much faster than the decay rate $\sim g^2m_\phi^3$, since at early times the temperature exceeds $m_\phi$. However, this rate is thermally averaged over ALP energies with a thermal distribution. 
If this rate would apply to the thermalization of the condensate, i.e.~of the zero mode, the condensate would perish if $g^2T^3$ ever exceeded the expansion rate $H\sim T^2/m_{\rm Pl}$. The higher temperatures would be the most relevant and the condensate survives if $g\lesssim 1/\sqrt{T_R m_{\rm Pl}}$. 
However, let us convince ourselves that this is not the case. 

First, note that the inverse-Primakoff process $\phi +e^\pm\to \gamma +e^\pm$ is exponentially suppressed at high temperatures because the energy of the incoming electron $E$ has to be enough to produce a photon, which at finite temperature has a non-zero mass $\sim eT$. This threshold implies $2 m_\phi E>m^2_\gamma$ or $E/T>e^2 T/m_\phi$ which can be huge if $m_\phi$ 
is tiny\footnote{One can check that after electron-positron annihilation, when the formulas we have used are not valid anymore, the evaporation is still slow.}.  Since the abundance of these electrons will be exponentially suppressed so would be the rate. 

This is however not a showstopper, since the threshold can be easily overcome by considering an additional photon in the initial state, i.e.~$\gamma+\phi +e^\pm\to \gamma +e^\pm $. 
However, these and similar processes are also suppressed at high temperatures $T\gg m_\phi$.
The reason is that the two-photon coupling involves the derivative of the axion field (since $F_{\mu\nu}\widetilde F^{\mu\nu}$ is a total derivative) whose only component is the time component $\partial_0 \phi\sim m_\phi \phi$ (after $t_1$) and therefore all amplitudes involving the zero mode ALP absorption are necessarily proportional to $m_\phi$ and absorption probabilities to $m_\phi^2$. 
To provide a more complete example, we can compare the rate of Compton scattering of a photon of energy $\omega$, denoted by $\Gamma_C(\omega)$, with the rate of the process $\gamma(\omega)+\phi +e^\pm\to \gamma +e^\pm$. In the limit of very small $m_\phi$ the ALP absorption and the virtual photon scattering factorize and the differential ALP absorption rate due to photons of energy $\omega$ is
\bea
d\Gamma_{\phi C} &=& \frac{1}{2 m_\phi}\Gamma_C(\omega)|{\cal M}(\gamma\phi\to \gamma^*)|^2
\frac{1}{(2\omega m_\phi)^2 + (\omega\Gamma_C(\omega))^2} d n_{\gamma}(\omega)\\
&=&  g^2 m_\phi \beta^2
\frac{\omega^2\Gamma_C(\omega)}{(2\omega m_\phi)^2 + (\omega\Gamma_C(\omega))^2}  d n_{\gamma}(\omega)
\eea
where $d n_{\gamma}(\omega)$ is the density of photons with energy $\omega$ and $\beta \sim 1$ is the photon velocity. We have included $\Gamma_C$ as the imaginary part of the propagator to account for the finite photon lifetime in the thermal bath. 
For the rate of ALP absorption we have to integrate over the initial photon energies thermally distributed. 
In the $m_\phi\to 0$ limit it is 
\be
\Gamma_{\phi C} \sim g^2 T^3 \frac{m_\phi}{\langle \Gamma_C\rangle}
\ee
where $\langle \Gamma_C\rangle$ denotes a certain thermal average, expected to be $\sim \alpha^2T $ bearing phase space factors. 
In practice this is a very large suppression factor with respect to the naive thermalization rate $g^2T^3$, which leaves the parameter space shown in Fig.~\ref{fig:ALPDM} untouched. 
We believe that a similar kind of reasoning can be applied to other ALP couplings of derivative type, like to fermions\footnote{It seems that this argument was not taken into account in the axion case~\cite{Turner:1986tb}. However, there it has no phenomenological consequences.}. 
We plan to expand on these arguments at length in a forthcoming publication. 

{If large scale magnetic fields are generated in the early universe, for instance during the electroweak phase transition, they can trigger ALP-photon oscillations. The ALP condensate then oscillates into a large scale coherent electric field which can be easily damped due to the high conductivity of the primordial plasma~\cite{Ahonen:1995ky}. However, the high conductivity strongly suppresses the ALP-photon mixing and in practice the required magnetic fields are unreasonably large. 
For instance, if we assume magnetic fields are frozen into the medium, their strength redshifts as $B=B_0(1+z)^2=B_0(a_0/a)^2$ and requiring no damping of the condensate produces the constraint
\bb
\label{eq:ALP_res_bound}
\(\frac{g}{10^{-10}\, {\rm GeV}^{-1}}\)^2\(\frac{B_0}{3{\rm{n}} {\rm G}}\)^2\frac{T_B}{{10^9\, \rm GeV}} \lesssim 1
\ee
where $T_B$ is the temperature of the universe when magnetic fields form. This damping would have implications in the ALP parameter space Fig.~\ref{fig:ALPDM} if strong primordial magnetic fields of very early origin are eventually discovered and conversely, if WISPy ALP DM is established by direct detection experiments, one could use this bound to constrain the existence of magnetic fields in the very early universe. 
If the magnetic fields are produced during the electroweak phase transition, the plausible constrains are not very promising. 
One would then need to have primordial fields of the order of $B\sim T^2$ (close to equipartition with radiation), which means a value $B_0\sim 3\mu$G today, for them to have consequences for ALPs which are not already excluded by stellar evolution, i.e. for $g<10^{-10}$ GeV$^{-1}$. These fields are very close to the exclusion limits for primordial magnetic fields (if not excluded) and they would imply distortions of the CMB spectrum for sub-eV mass ALPs, see~\cite{Mirizzi:2009nq} and references therein. 

Note that the {constraint \eqref{eq:ALP_res_bound} is} independent of the ALP mass. This happens for relatively large values of $T_B$ but if the primordial fields are created relatively late, a resonance in the ALP-photon oscillations (not considered in~\cite{Ahonen:1995ky}) will dominate the ALP condensate evaporation, and this does depend on the ALP mass. This resonance is very sharp in time and only depends on the size of the magnetic fields at that particular time, and therefore is independent of the very early universe.  
We postpone the quantitative aspects to the next section, devoted to  hidden photons, because HP-photon oscillations are the most relevant HP evaporation mechanism and the required formalism is essentially the same.
The resulting bound is displayed {in Fig.~\ref{fig:ALP_Bresonance}.}

\begin{figure}[t] 
\centering   \includegraphics[width=4in]{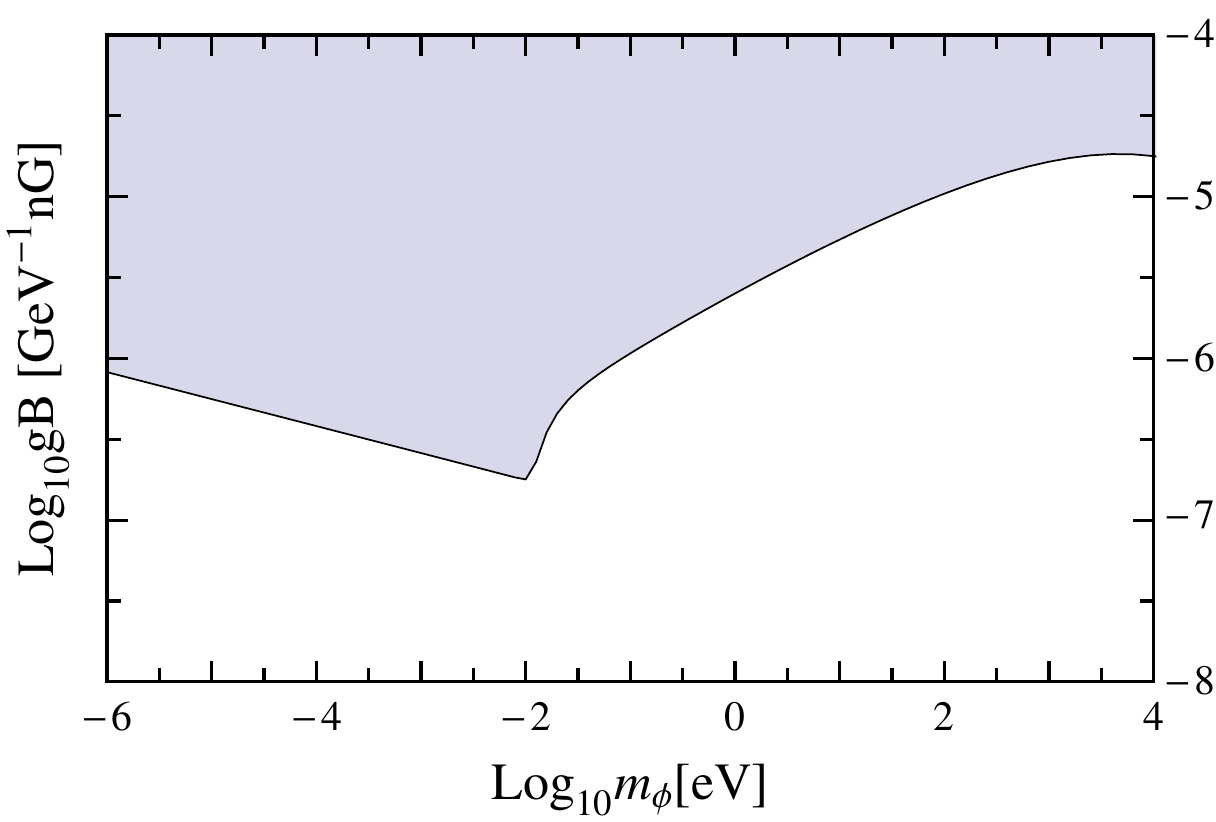} 
\caption{{Region that would be affected by resonant ALP-photon oscillations leading to the evaporation of the ALP condensate in the case where primordial magnetic fields exist and they evolve with redshift as $B=B_0(1+z)^2$. }}
\label{fig:ALP_Bresonance}
\end{figure}
 }

\subsection{Thermal population of ALPs}

The Primakoff process is able to create a thermal ALP population if $\Gamma_{\phi q^\pm}$ exceeds the expansion rate at some point in the history of the universe. This population competes with the condensate to form DM. 
The phenomenological implications of this population have been recently reviewed in~\cite{arXiv:1110.2895} (see also~\cite{Masso:2004cv}). 
In particular, thermal ALP CDM exceeds the measured value Eq.~\eqref{eq:observedDM} unless 
\be
m_\phi < 154\,  {\rm eV} \left(\frac{106.75}{g_*(T_d)}\right),  
\ee
where $T_d$ is the temperature at which $\Gamma_{\phi q^\pm}\sim H$ where the ALP bath decouples from the SM bath. This bound assumes that 
\begin{equation}
T_d\sim 10^4\,{\rm GeV}\left(\frac{10^{-10}\,{\rm GeV}^{-1}}{g}\right)^2
\end{equation}
is somewhat smaller than the reheating temperature.
In the large mass region this puts a quite stringent limit on ALP CDM. 
However, it can be circumvented in scenarios with a low reheating scale and we therefore do not include it in Fig.~\ref{fig:ALPDM}.

\subsection{Detecting photons from ALP decay}

\begin{figure}[tbp] 
   \centering
   \includegraphics[width=4in,trim= 0 0 -0.2cm 0]{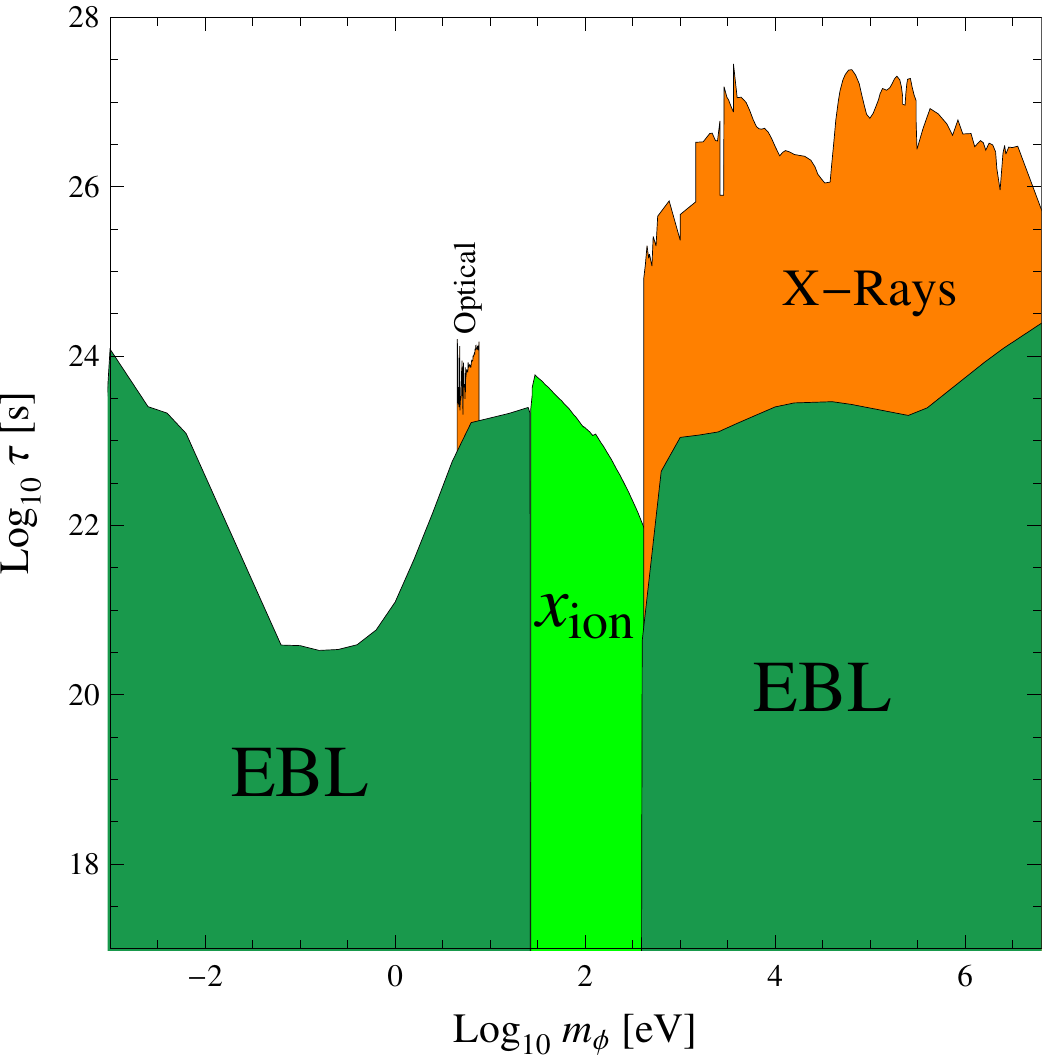}   
   \caption{Exclusion bounds on axion-like particles from relic photons in the mass--lifetime parameter space.}
   \label{fig:relicphotons}
\end{figure} 

Even if the ALP lifetime is longer than the age of the universe some ALP decays inevitably happen and the resulting monochromatic photons can signal the existence of ALP CDM.  
We show the bounds coming from relic photon detection in Fig.~\ref{fig:ALPDM} and more specifically in Fig.~\ref{fig:relicphotons}, where they are plotted in the $m_\phi-\tau$ parameter space (where $\tau$ is the ALP lifetime).

To obtain these bounds we have compared the extragalactic background light (EBL) spectrum from Ref.~\cite{Overduin:2004sz} with the estimated flux of photons due to the decay of the ALP~\cite{Masso:1997ru,Masso:1999wj}, 
\begin{align}\label{eq:fluxMD}
\frac{dF_E}{dEd\Omega}&=\frac{1}{2\pi}\frac{\Gamma_\gamma}{H(z)}\frac{\rho_\phi(z)/m_\phi}{(1+z)^3}=\nonumber \\
&\simeq \frac{\rho_{\phi 0}/m_\phi}{2\pi\tau H_0}\left(\frac{E_0}{m_\phi/2}\right)^{3/2}\exp\left(-\frac{t_0}{\tau}\left(\frac{E_0}{m_\phi/2}\right)^{3/2}\right)
\theta\left(m_{\phi}-2E_{0}\right)
\end{align}
where we wrote all present quantities with a subscript ``0'' and the decay-redshift as $1+z=(m_\phi/2)/E_0$, with $E_0$ being the photon energy today.

The EBL spectrum shows no particular features. Thus it can be used to exclude any ALP which would provide too strong signal. From the comparison we exclude the portion of parameter space labelled ``EBL'' in  
Figs.~\ref{fig:ALPDM} and~\ref{fig:relicphotons}. 

Galaxies, being denser, should provide an enhanced signal for decaying ALPs. 
Again, if the decay rate of ALP CDM is high enough, we should be able to detect a spectral line whose energy is $E=m_\phi/2$.
Axions in the visible part of the spectrum~\cite{Ressell:1991zv,Bershady:1990sw,Grin:2006aw} and sterile neutrinos in X-ray spectra~\cite{Boyarsky:2006fg,Boyarsky:2009ix} have already been searched using this technique. 
Following these references and conveniently rescaling their results for ALPs, we obtain the exclusion bounds in Figs.~\ref{fig:ALPDM} and~\ref{fig:relicphotons} labelled ``Optical'' and ``X-Rays'', respectively. 
{The bounds from searches of gamma-ray lines in the data from FERMI~\cite{Vertongen:2011mu} have been discussed in~\cite{arXiv:1110.2895}.  
Unfortunately, the very suggestive yet tentative claim of a detection at $E_0\simeq 130$ GeV~\cite{Weniger:2012tx}, can only be attributed to annihilating DM and not to decaying DM.}

The search for ALP decay photons is blinded in the ultraviolet range by the humongous opacity of the atmosphere. However, a substantial amount of ionizing radiation at high redshifts can strongly affect the reionization history of the universe. 
The optical depth of reionization has been estimated by WMAP7 to be $0.088\pm0.015$, out of which nearly a factor $0.04-0.05$ can be attributed to a fully ionized universe up to redshift $\sim 6$ as supported by the absence of Ly-$\alpha$ features in quasar spectra. 
The origin of the remaining part of the optical depth is still matter of debate, and thus we require the ionization caused by decaying ALPs to be less than what this fraction would require.
Assuming one ionization per ALP decay photon, we have computed the reionizaton history using RECFAST~\cite{astro-ph/9909275} 
and obtained the modified optical depth from $z=6$ to $z=100$, $\tau_6$. 
In the region labelled ``$x_{\rm ion}$'' in Figs.~\ref{fig:ALPDM} and~\ref{fig:relicphotons}, ALP decays produce too early reionization and exceed the measured $\tau_6\simeq 0.04$ by 1 standard deviation. 
This is a conservative bound. If we assume that the full energy of the photon can be converted into ionization this bound strengthens increasingly with mass up to one order of magnitude at the largest masses for which ionization is effective, $m_\phi\lesssim 300$ eV. 

\subsection{{Other indirect observational constraints on ALPs}}

{The propagation of photons and cosmic rays in today's ALP background might have potentially detectable consequences. 
The ALP background implies time-dependent birefringence which produces rotation of the polarization plane of photons propagating across the universe~\cite{Harari:1992ea}. For simplicity, let us consider the case in which the ALP field is homogenous and the frequency of light is bigger than the ALP mass, $\omega\gg m_\phi$. Then the rotation of the polarization plane in the WKB approximation is independent of $\omega$ and given by the line-of-sight integral
\bea
\nonumber
\Delta \varphi &\simeq& g \int dl \dot{\phi}\sim \frac{g \sqrt{\rho/2}}{m_\phi} \(\cos (m_\phi l+\beta)-\cos (\beta)\) \\
\label{eq:tralara}
&\simeq& 2\times 10^{-25} \frac{g}{{10^{-10}\rm GeV}^{-1}} \frac{{\rm eV}}{m_\phi}\(\cos (m_\phi l+\beta)-\cos (\beta)\)
\eea
where $\beta$ is an unknown {phase and we have neglected the expansion of the universe (including the expansion is straightforward)}. 
Clearly, {this effect is ${\cal O}(1)$ only for extremely small masses, and this only if we observe 
objects at distances $\gg 1/m_{\phi}$.} {For instance, the bounds implied by the non-observation of such rotation in the emissions of AGNs~\cite{Horns:2012pp} seem to be irrelevant. 
Indeed, recalling that todays ALP density, and therefore its field, is bounded by (\ref{eq:ALPDMbound})
and using $\phi_1={\cal N}\theta_1\alpha/2\pi g$ with ${\cal N}=1$ and $\theta_{1}<\pi$ one finds 
\bb
\label{constra}
\Delta \varphi < {10^{-8}}\sqrt{\frac{1.8\times 10^{-27}\,\rm eV}{m_\phi}}\(\cos (m_\phi l+\beta)-\cos (\beta)\).
\ee 
This shows that ALP DM can} produce only minute polarization changes\footnote{{This rotation is proportional to the amplitude of the sidebands looked for in the experiment proposed in~\cite{Melissinos:2008vn} to search for meV mass axions or ALPs. 
{Therefore, Eq.~\eqref{constra} seems to make their detection quite challenging}.}}. 
{In \eqref{eq:tralara} and \eqref{constra} we have used the average value of DM which is much smaller than the values currently present in the galaxy or in the universe after recombination, which are bigger by factors  $\sim 10^5$ and $10^9$, respectively. Unfortunately, this does not change our conclusions. Even going to the largest possible distances, i.e. looking at the CMB, the effects have been found to be small in the range of masses relevant for ALP DM~\cite{arXiv:0905.4720}.}
Other consequences, like the existence of gaps in the photon dispersion relation at frequencies close to integer multiples of $m_\phi$ have been discussed in the literature~\cite{Espriu:2011vj} and found to be not promising for the detection of ALPs. 

Another possible signature would be the radiation generated by cosmic rays propagating in the time-varying ALP background. 
Low energy radiation was considered in~\cite{Espriu:2010ip}  and found to be many orders of magnitude below the backgrounds. 
At high energies, the inverse Primakoff process $q+\phi\to q+\gamma$ could in principle contribute to the relatively small gamma-ray backgrounds, however the contribution arising from of {inverse Compton scattering of cosmic rays off the} CMB photons $q+\gamma_{\rm CMB}\to q+\gamma$ dominates because of the much larger cross section.  }

\subsection{Direct experimental and observational constraints on ALPs}

So far we have focussed on the astrophysical and cosmological constraints specific to ALP CDM.
However, in Fig.~\ref{fig:ALPDM} we also show a variety of constraints that arise from the properties of ALPs alone without them needing to 
form all or part of the DM.
In the mass region shown in Fig.~\ref{fig:ALPDM} three constraints are most important.
The constraint labelled ``ALPS'' is the result of the light-shining-through-walls experiment\footnote{In such an experiment strong magnetic fields are used to induce photon-ALP oscillations in incoming laser light. The very weakly coupled ALP state can then pass through a
wall. On the other side the ALPs can then oscillate back into photons which can be detected. See~\cite{Redondo:2010dp} and references therein. For HPs the same technique works but no magnets are required.} ALPS~\cite{Ehret:2010mh}.
As we can see this experiment does not yet test the ALP CDM region. However, in the near future significant improvements by several orders
of magnitude are expected by enhancing the signal with resonant cavities~\cite{Hoogeveen:1990vq,Sikivie:2007qm}. Moreover improvements
are expected from upcoming experiments in the microwave regime~\cite{Hoogeveen:1992uk,Jaeckel:2007ch,Caspers:2009cj}

The ``CAST+SUMICO'' constraint arises from the helioscopes\footnote{Helioscopes use the same idea as light-shining-through walls experiments. 
The ALPs are however produced inside the sun from photons interacting with the electromagnetic fields of electrons and ions in the plasma.}~\cite{Sikivie:1983ip} CAST~\cite{Andriamonje:2007ew,Arik:2008mq} and SUMICO~\cite{Inoue:2008zp}. As we can see these experiments already exclude sizable regions of the ALP CDM parameter space. Also for these experiments significant improvements are expected in the future. In particular if a next generation axion helioscope such as IAXO~\cite{Irastorza:2011gs} is realized.

Finally, the bound ``HB'' arises from comparing the observed cooling rate of horizontal branch stars with the expected rate. This places strong bounds on
additional energy losses caused by a production of ALPs in the star's core~\cite{Raffelt:1985nk,Raffelt:1996wa}. 
These bounds are currently the strongest and probe the ALP DM region. However, as these bounds are limited by astrophysical uncertainties
we expect that the more controlled experiments discussed above will overtake them in the not too distant future. 

As already alluded none of these experiments make use of ALPs being DM. This makes them particularly model independent, but also ignores
a potential plentiful source of ALPs. We will return to haloscopes which indeed exploit this source in Sect.~\ref{sec:direct}.

\section{Hidden photons}\label{sec:HPs}

In a recent 
article~\cite{arXiv:1105.2812}, Nelson and Scholtz {have considered the possibility that the misalignment mechanism could also be applied to generate a population of hidden photons (HPs), an Abelian} gauge boson under which SM particles are uncharged. The Lagrangian is   
\bb 
\label{hidden}
\mathcal L= -\frac{1}4X_{\mu\nu}X^{\mu\nu}+\frac{m_{\gamma'}^2}2 X_\mu X^\mu + {\cal L}_{\rm grav}+{\cal L}_I,
\ee
where $X^{\mu}$ is the HP gauge field and $X^{\mu\nu}$ its field strength. {Moreover ${\cal L}_I$ contains the interactions with the Standard Model particles and ${\cal L}_{\rm grav}$ specifies potential non-minimal gravitational couplings discussed below.}    
The HP mass might result from the Higgs or St\"uckelberg mechanisms. In the first case, we have to worry when the phase transition happens and we might have a similar scenario to the one sketched in the previous section for Nambu-Goldstone bosons. Also a Higgs particle appears in the spectrum, with mass $\sim \sqrt{\lambda} m_{\gamma'}/g_h$ 
where $g_h$ is the hidden sector gauge coupling and $\lambda$ the Higgs self-coupling. Even if we take $g_h$ to be really small, the Higgs particle phenomenology tightly constrains this scenario, especially for the sub-eV values of $m_{\gamma'}$ we explore~\cite{arXiv:0807.4143}.  As in the original proposal~\cite{arXiv:1105.2812}, we focus therefore on the St\"uckelberg case, which occurs naturally 
in large volume string {compactifications~\cite{Burgess:2008ri,Goodsell:2009xc,Cicoli:2011yh}. 
In this case, there is no SSB phase transition.}

{
Let us briefly discuss the evolution of the HP in an expanding Universe\footnote{{We would like to thank Valery Rubakov and Christof Wetterich for 
noticing an error in the treatment of the cosmological evolution of a vector field (which is also present in Ref.~\cite{arXiv:1105.2812}).}}. 
For reasons that will become clear at the end of this discussion we also include a non-minimal coupling to gravity of the form\footnote{{We use a 
coordinates such that $ds^2=dt^2-a^{2}(t)dx^{2}_{i}$, i.e. the metric is $g_{\mu\nu}={\rm diag}(1,-a^{2}(t),-a^{2}(t),-a^{2}(t))$. Moreover, the
gravitational part of the Lagrangian is ${\cal L}_{\rm GR}=-R/(16\pi G_{N})$.}}
\begin{equation}
{\cal L}_{\rm grav}=\frac{\kappa}{12}R X_{\mu}X^{\mu}.
\end{equation}

For simplicity let us focus on the homogeneous solution, 
$\partial_{i}X_{\mu}=0$.
The equation of motion then enforces $X_{0}=0$.
As explained in~\cite{Golovnev:2008cf} the invariant $X^{\mu}X_{\mu}=-1/a^{2}(t) X_{i}X_{i}$ is a coordinate independent measure for the size of the vector and it is convenient to introduce $\bar{X}_{i}=X_{i}/a(t)$. 
Using this the equation of motion is,
\begin{equation}
\label{vecevolution}
\ddot{\bar{X}}_{i}+3 H\dot{\bar{X}}_{i}+\left(m^2_{\gamma^{\prime}}+(1-\kappa)(\dot{H}+2H^2)\right)\bar{X}_{i}=0.
\end{equation}
The energy density is
\begin{equation}
\label{vecdensity}
\rho(t)=T^{0}_{0}=\frac{1}{2}\left(\dot{\bar{X}}_{i}\dot{\bar{X}}_{i}+m^{2}_{\gamma^{\prime}}\bar{X}_{i}\bar{X}_{i}+(1-\kappa)H^2 \bar{X}_{i}\bar{X}_{i}+2(1-\kappa) H\dot{\bar{X}}_{i}\bar{X}_{i}\right).
\end{equation}

For $H\ll m_{\gamma^{\prime}}$ and $\dot{H}\ll m^{2}_{\gamma^{\prime}}$ the expressions~\eqref{vecevolution} and \eqref{vecdensity} reduce
to the same form as Eqs.~\eqref{eq:condensev}  and \eqref{scalardensity}, independent of the value of $\kappa$.
In consequence, the same approximate solution Eq.~\eqref{eq:CCDM} holds for the HP case. 
In particular for $m_{\gamma^{\prime}}={\rm const}$  the energy density behaves (for any value of $\kappa$) just like that of non-relativistic matter $\rho(t)\sim 1/a^{3}(t)$.

In the following we will derive constraints that arise from the evolution in this phase of the evolution. When the expansion rate is slow one can check
that $\bar{X}_{i}$ actually is the properly normalized field. Therefore, from the following subsection on, we will simply drop the bar.  

Let us now have very brief a look at the initial conditions and the evolution before oscillations begin.
In contrast to the ALP case, there is no natural value for the initial value of the amplitude\footnote{If the hidden photon mass arises from a Higgs mechanism one may wonder if there is no limitation on the field value from the fact that one component of the field arises from the ``eaten'' Goldstone boson which has a limited field range, similar to the situation we discussed for ALPs. However, in effect the extra field component of the gauge field corresponds to a derivative of the Goldstone boson, which is not bounded.
Similar reasoning can be applied to the St\"uckelberg case.},  $\bar{X}_{1,i}$.  
Therefore choosing a suitable value for $\bar{X}_{1,i}$ any HP model can provide DM, i.e.~nature might have tuned the value of $\bar{X}_{1,i}$ to fit the observed value Eq.~\eqref{eq:observedDM}.
However, one might want to be a bit more ambitious and also consider the evolution before $t_{1}$. 
Here, we will only consider the simplest case, $\kappa=1$. Then Eqs.~\eqref{vecevolution} and \eqref{vecdensity} reduce to the same form as in the scalar case~\cite{Golovnev:2008cf}.  Accordingly, for $H\gg m_{\gamma^{\prime}}$ both the field $\bar{X}_{i}$ and the energy density are approximately constant, and we can think of these values being stuck there from some time at the beginning of, or even before, inflation. 
}

One further comment is in order. The direction of the $\mathbf{X}$ field remains unchanged for most of the universe history. However, it is conceivable that it changes during the process of structure formation where inhomogeneities in the gravitational potential grow relevant and Eq.~\eqref{eq:condensev} is no longer valid.  
This has important consequences for direct detection, that will be discussed in Sect.~\ref{sec:direct}.  

\subsection{Survival of the condensate}

Under the assumption that SM particles are uncharged under the hidden photon gauge group,
the dominant interaction between the visible and the hidden sector is through gauge kinetic mixing between  
photons and hidden photons~\cite{Holdom:1985ag},
\bb 
\label{HPlagrangian}
\mathcal L= -\frac{1}4F_{\mu\nu}F^{\mu\nu}-\frac{1}4X_{\mu\nu}X^{\mu\nu}+\frac{m_{\gamma'}^2}2 X_\mu X^\mu-\frac{\chi}2 F_{\mu\nu} X^{\mu\nu}+ J^\mu A_\mu,
\ee
where $A_\mu$ is the photon field and $F_{\mu\nu}$ the corresponding field strength, and $J^\mu$ is the current
of electrically charged matter. Kinetic mixing is generated at one-loop by the exchange of heavy messengers that couple both
to the ordinary photon as well as to the hidden photon, its natural value therefore being determined by 
the visible and hidden gauge couplings via $\chi\sim e g_h/(16\pi^2)$. In field theory and in compactifications of heterotic 
string theory, the hidden gauge coupling is of order one and thus $\chi\sim 10^{-3}$~\cite{Dienes:1996zr,Goodsell:2011wn}. 
In large volume string compactifications,  
the hidden gauge coupling $g_h$ can be very small and there is no clear minimum for $\chi$: 
values in the $10^{-12}$---$10^{-3}$ range have been predicted in the literature~\cite{Goodsell:2009xc,Cicoli:2011yh}.  

{For a St\"uckelberg mass the same string scenarios typically prefer values $\gtrsim 10^{-4}\,{\rm eV}$.
Nevertheless, to be as inclusive and model-independent as possible we will take both $\chi$ and $m_{\gamma^{\prime}}$ as free parameters in our phenomenological study.}

By means of the re-definition $A_\mu = \tilde A_\mu-\chi \tilde X_\mu$, $X_\mu = \tilde X_\mu$ we can identify the propagation basis \emph{in vacuum} ($\tilde A_\mu, \tilde X_\mu$), where the kinetic mixing has been removed and the Lagrangian looks like
\bb \mathcal{ \tilde L}=  -\frac{1}4\tilde F_{\mu \nu}\tilde F^{\mu \nu}-\frac{1}4 \tilde X_{\mu\nu}\tilde X^{\mu\nu}+\frac{m_{\gamma'}^2}2 \tilde  X_\mu \tilde X^\mu+ J^\mu \left(\tilde A_\mu-\chi \tilde X_\mu \right). 
\label{lagrangian}
\ee

The universe is not empty though. The interactions of photons with the charged particles in the primordial plasma induce refraction and absorption. This can be described with an effective photon mass squared $M^2\equiv m_\gamma^2+i\omega\Gamma$, where both the plasma mass 
$m_\gamma$ and the interaction rate $\Gamma$ only depend\footnote{In an isotropic homogeneous universe. } on the photon frequency, $\omega$, and the modulus of the wavenumber, $k$~\cite{Redondo:2008ec}. 
The corresponding effective term in the Lagrangian can be written as ${\cal L}=M^2A_\mu A^\mu/2$. 
This is non diagonal in the $\{\tilde A,\tilde X\}$ basis and crucially determines the propagation eigenstates and their eigenvalues~\cite{Redondo:2008ec,Redondo:2008aa,Jaeckel:2008fi,Mirizzi:2009iz}. 

$M^2$ {grows with  the charge density $n_Q$ and the temperature and therefore} varies significantly during the history of the Universe. After inflation we can assume {the universe has been efficiently emptied} and thus $M^2=0$. {During reheating, $n_Q$ and $T$ grow very rapidly} up to extremely large values and then decrease with smooth power laws until today, where we shall assume them to be zero again.   
Forgetting pre-/reheating times for a moment, at sufficiently early times $M^2$ is the largest scale in the  problem. Neglecting $H$, the e.o.m.~for the photon $A=\widetilde A-\chi \widetilde X$ is that of a damped oscillator, while the orthogonal combination $\widetilde X+\chi\widetilde A$ is (almost) a free field with a mass $\simeq m_{\gamma'}$. The latter is the combination which can form a condensate. 

Today $M^2$ tends to $0$, so $\widetilde A$ and $\widetilde X$ are the propagation eigenstates, their e.o.m.~decouple and both follow the solution outlined in Sect.~2.  
Therefore, the condensate has to follow a trajectory in field space from $\widetilde X+\chi\widetilde A$ to $\widetilde X$ without losing too much amplitude. 

Assuming that the universe expands slowly compared to the microscopic timescales of photon absorption ($d\Gamma /dt \ll \Gamma^2$, or $\Gamma\gg H$) the two eigenstates adiabatically follow their static solutions. 
The adiabatic propagation eigenstates are given by an effective mixing angle, $\chi_{\rm eff}$, the angle  between the sterile state 
($\widetilde X+\chi \widetilde A$) and the condensate. The eigenstates decay with rates $\Gamma_1=(1-\chi^2_{\rm eff})\Gamma$ and $\Gamma_2=\chi^2_{\rm eff}\Gamma$. At the lowest order in $\chi$, the effective mixing angle can be approximated by 
\bb 
\chi^2_{\rm eff}\simeq  \frac{\chi^2 m_{\gamma'}^4}{\left(m_\gamma^2-m_{\gamma'}^2\right)^2+ \A ^4} .
\label{chieff}
\ee 
where $\A^2={\max}\{\chi m_{\gamma'}^2,m_{\gamma'}\Gamma\}$.  

Before recombination $m_\gamma^2$ is positive so there might be a moment where the mixing is resonant.
Characterizing it by the temperature of the universe at the time of the resonance $m^2_\gamma(T_{\rm res})=m^2_{\gamma'}$  
we can distinguish three regimes
\begin{itemize}
\item {\it{ At high temperatures $T\gg T_{\rm res}$:}} $m_\gamma\gg m_{\gamma'}$ and the mixing is suppressed as $\chi_{\rm eff} \simeq \chi m_{\gamma'}^2/{\sqrt{m_{\gamma}^4 +\mu^4}} \ll \chi$.
\item {\it{At low temperatures $T\ll T_{\rm res}$:}} $m_\gamma\ll m_{\gamma'}$ and we recover the vacuum mixing parameter $\chi_{\rm eff}\simeq \chi$ {(since usually $\mu^4<m_{\gamma'}^4$)}.
\item {\it{At resonance $T=T_{\rm res}$:}} the mixing is significantly enhanced. If $\chi m_{\gamma} > \Gamma$ the mixing angle is maximal and one 
can have a resonant transition similar to the MSW effect in neutrinos. For small $\chi$ the mixing still is enhanced with respect to the vacuum case by a factor $m_\gamma/\Gamma\gg 1$. 
\end{itemize}

Thus, in general the condensate does not move smoothly in field space from $\widetilde X+\chi\widetilde A$ to $\widetilde X$. It overshoots $\widetilde X$ during the resonance only to then return to $\widetilde X$. 

\begin{figure}[t]
\centering
\includegraphics[scale=1.0]{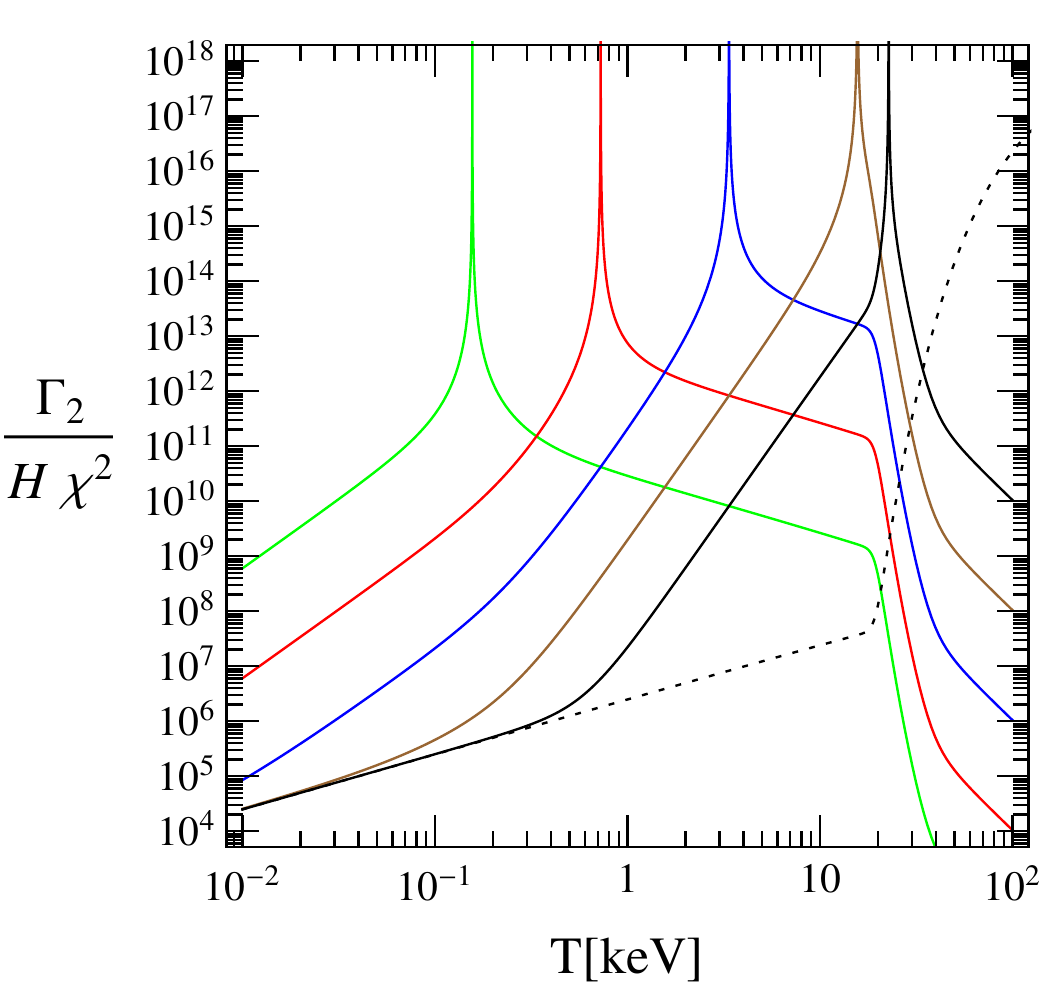}
\caption{The decay rate of the HP condensate,  $\Gamma_{2}$, normalized to $H\chi^2$, as a function of temperature, for different HP masses. {The curves, ordered by their resonance from left to right, correspond to $\muu=10^{-5},10^{-4},10^{-3},10^{-2}$ and 0.1 eV. The dashed line corresponds to the expression used in Ref.~\cite{arXiv:1105.2812}.}}
\label{fig:iso}
\end{figure}

Let us first discuss the case of over-damped oscillations during the resonance, i.e. 
$\chi^2_{\rm eff}\sim \chi^2 m^{2}_{\gamma^{\prime}}/\Gamma^{2}$. 
The amplitude of the HP condensate decreases by the ratio
\bb
\label{ratio}
\frac{X_{2,\rm today}}{X_{2,\rm initial}} = \(\frac{a_{\rm initial}}{a_{\rm today}}\)^{3/2}\exp\(-\frac{1}{2}\int_{t_{\rm initial}}^{t_{\rm today}}dt \Gamma_2 \). 
\ee
Approximating $T/T_0=a_0/a$, we can write $\Gamma_2 dt$ as $\Gamma_2/H d\log T$. 
The function $\Gamma_2/H\chi^2=\Gamma\chi^2_{\rm eff}/H\chi^2$ is plotted for different HP masses in Fig.~\ref{fig:iso}. 
{In the plot we have used 
\bb
M^2= i \omega {\bf \sigma} = \omega_{\rm P}^2\frac{(\omega \tau)^2+ i \omega \tau}{1+(\omega \tau)^2}+ i\omega \Gamma_{\rm Th}
\ee
where $\omega=m_{\gamma'}$ is understood, $\omega_{\rm P}$ is the plasma frequency, $\Gamma_{\rm Th}=\sigma_{\rm Th}n_e$ is the absorption due to Thomson scattering off the electrons and $\tau$ the characteristic time between electron collisions, which sets the friction time scale\footnote{The zero-frequency conductivity $\sigma_0=\omega_{\rm P}^2 \tau$ is normally used instead of $\tau$. We have followed the approximations for $\sigma_0$ that can be found in~\cite{Ahonen:1995ky} and supplemented them in the non-relativistic electron case with Thomson friction $\sigma_0\approx\frac{45}{2 \pi^2 \alpha} \frac{n_e m_e^2}{T^4}$ where $\alpha$ is the fine-structure constant, and $n_e,m_e$ the electron density and mass.}.}

As expected the decay rate is heavily enhanced at $T_{\rm res}$, and rapidly drops at higher temperatures. 
The integral  can be well approximated by the contribution near the resonance. 
Using $m_\gamma^2=m_{\gamma'}^2+|dm_{\gamma'}^2/dT|_{\rm res}(T-T_{\rm res})$ and evaluating the other $T$-dependent quantities at $T_{\rm res}$ we have
\bb
\label{HPresult}
\frac{1}{2}\int_{T_{\rm today}}^{T_{\rm initial}}\frac{dT}{T} \frac{\Gamma_2}{H}\simeq   \chi^2 \frac{\pi}{2} \frac{m_{\gamma'}}{r H_{\rm res}}\equiv \frac{\tau_2}{2}
\ee
where $r=d\log m_\gamma^2/d\log T$ is an ${\cal O}(1)$ factor, see Fig.~2 of \cite{Redondo:2008ec}. Moreover, $\exp(-\tau_{2})$ gives the damping of the total energy density due to the resonance. 
Note that the result is independent of $\Gamma$. 

If the resonance is not over-damped, i.e. $\chi \muu > \Gamma$, the HP survival probability after the level-crossing can be approximated by the Landau-Zener expression used in~\cite{Mirizzi:2009iz}. The result coincides almost exactly with (\ref{ratio}), only requiring an extra factor of 2 in the argument of the exponential, i.e. $\tau_2\to 2\tau_2$.

The plasma effects {were} neglected in Ref.~\cite{arXiv:1105.2812} which forced the authors to conclude that the most relevant period for the evaporation of the condensate was at high temperatures  (the corresponding ``approximation'' is shown as a dotted line in Fig.~\ref{fig:iso}). As we have seen, at high temperatures the evaporation process is strongly suppressed by the effective mixing angle. 
The resonance dominates the condensate evaporation and for the sub-eV masses of main interest here happens at temperatures smaller than the electron mass~\cite{Jaeckel:2008fi}. 

{Before we continue, it is now time to describe the differences between HP-photon and ALP-photon oscillations in the presence of a background homogeneous magnetic field of primordial origin. 
The physics is essentially the same, except that for ALP-photon oscillations the mass mixing term in (\ref{chieff}) is $g B m_{\phi}$~\cite{Ahonen:1995ky} with $B$ the magnetic field strength, instead of $\chi \muu^2$. The important difference is that the value $B$ is expected to change during the evolution of the universe. The most relevant effects appear when magnetic fields are frozen to the medium, in which case $B=B_0(a_0/a)^2\simeq B_0(T/T_0)^2$. In this case, the {integral~\eqref{HPresult} tends} to be dominated by the highest temperatures, which in this case are the temperatures at which the magnetic fields are created $T_B$
\bb
\left.\frac{\tau_2}{2}\right|_{\rm ALP}\approx 
\(\frac{g}{10^{-10}\, {\rm GeV}^{-1}}\)^2\(\frac{B_0}{3{\rm n} {\rm G}}\)^2\frac{T_B}{{10^9\, \rm GeV}} . 
\ee
Imposing $\tau_2<1$ leads to the bound (\ref{eq:ALP_res_bound}). 
The resonance contribution can be computed, in full analogy to (\ref{HPresult}). We obtain 
\bb
\left.\frac{\tau_2}{2}\right|_{\rm ALP}\approx 
\frac{\pi}{2} \frac{g^2 B_{\rm res}^2}{r m_{\phi} H_{\rm res}}. 
\ee
}

\begin{figure}[tbp]
   \centering
   \includegraphics[width=15cm]{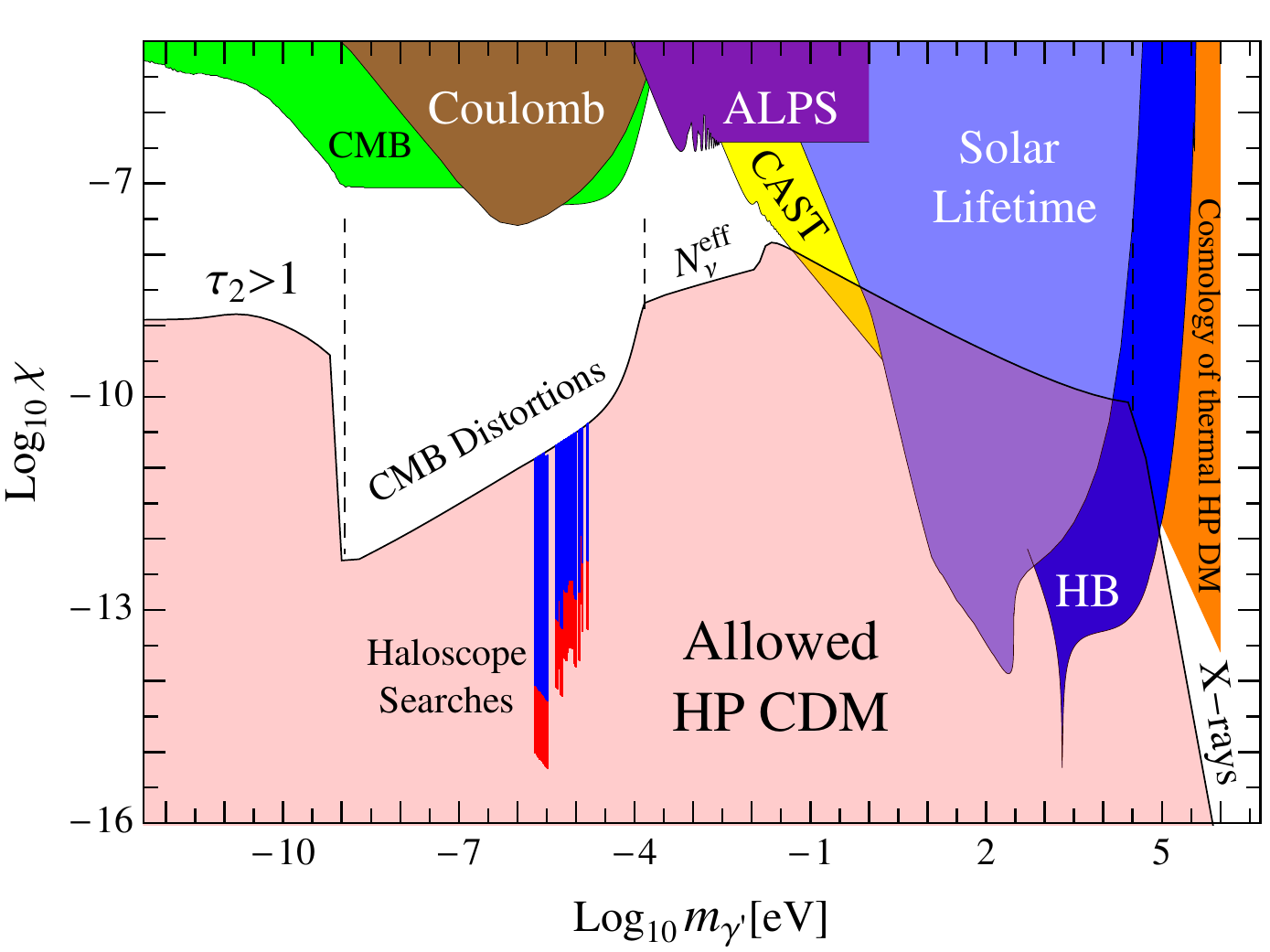} 
   \caption{Allowed parameter space for hidden photon cold dark matter (HP CDM) (for details see text).  The exclusion regions labelled ``Coulomb'', ``CMB'', ``ALPS'', ``CAST'' and ``Solar  Lifetime'' arise from experiments and astrophysical observations that do not require HP dark matter (for a review see~\cite{Jaeckel:2010ni}). 
We also show constraints on the ``cosmology of a thermal HP DM''. 
Note that only constraints on HPs with masses below twice the electron mass are shown since otherwise the cosmological stability condition requires unreasonably small values of the kinetic mixing, $\chi$.  The four constraints that bound the allowed region from above, ``$\tau_2>$1'', ``CMB distortions'', ``$N_\nu^{\rm eff}$'' and ``X-rays'' are described in the text.}
   \label{fig:HP}
\end{figure}

{Let us now come back to the discussion on HPs.} Since we can adjust the value of $X_{2,\rm initial}$ almost without restrictions\footnote{One may wish to impose that the initial field value is smaller than $m_{\rm Pl}$.  But this does not lead to any relevant additional constraints.}
 it seems that we can overcome any evaporation factor, even if it is enormously exponentially suppressed, and still have today HP CDM. 
There are, however, several limitations to this due to the fact that the evaporation process dumps some energy  into the photon bath, 
\bb
\Delta \rho=\rho_{\rm CDM}(e^{\tau_2}-1). 
\ee
This photon injection dilutes neutrinos and baryons with respect to photons, which can be constrained by the effective number of  relativistic neutrino species $N_\nu^{\rm eff}$ and by the successful agreement of CMB and big bang nucleonsynthesis (BBN) to estimate the baryon to photon ratio $\eta_B$.
Considering three massless neutrino species, a number of effective neutrinos smaller than 2.39 is excluded at 95\% C.L.~\cite{arXiv:1011.3694}, which translates into a photon temperature increase $T^\prime/T\sim1.06$. A similar value is obtained from BBN limits but the corresponding bound only applies to  injections between CMB and BBN times, while the $N_\nu^{\rm eff}$ one applies until the neutrino decoupling $T\sim 2$ MeV.  
If the injected photons thermalize, the photon temperature increases to
\bb
\label{tau2Temp}
T^\prime=\left( T_{\rm res}^4 +\frac{15}{\pi^2}\Delta\rho \right)^{1/4}\sim T_{\rm res}\left( 1+ 1.85 \frac{m_p}{T_{\rm res}} \eta_B \left(e^{\tau_2}-1 \right) \right)^{1/4},
\ee
where we used $\rho_{\rm CDM}\sim5 m_p \eta_B n_\gamma$, with $m_p$ being the proton mass and $n_\gamma$ the photon number density. 
We see from Eq.~\eqref{tau2Temp} that to obtain a significant increase in temperature one needs $\tau_2\gg-\ln(\eta_B)\sim 21$. 
Imposing\footnote{In this high $\tau_2$ regime, our formula~\eqref{HPresult} is not completely consistent. But, as long as $T'/T\sim 1$, 
it should provide a reasonable estimate.} $T^\prime/T_{\rm res}<1.06$, we can exclude the region above the curve labelled ``$N_\nu^{\rm eff}$'' in Fig.~\ref{fig:HP}. 

If the resonance happens below a critical temperature, the interactions of photons with the relic electrons and ions of the plasma might not be enough to fully recover a blackbody spectrum. The photons are initially injected at ultra-low energies, $\omega\sim \omega_{\rm Pl}\ll T$, at which inverse bremsstrahlung ($\gamma +e^- +p^+\to e^-+p^+$) and, to a lesser extent, inverse double Compton scattering ($\gamma+\gamma +e^-\to \gamma+e^-$) are always efficient in absorbing them. This results in the establishment of a blackbody distribution at a higher temperature than the initial, but only at low frequencies, where these processes are effective. The equilibration of the low energy and high energy parts of the spectrum can be achieved by a combination of Compton scattering (which is efficient in reshuffling photons up and down in energy) and the photon-number changing processes mentioned above which adjust the number density towards a blackbody.  
The injection of a relatively high number of low energy photons, $\delta n_\gamma/n_\gamma\gg \delta \rho_\gamma/\rho_\gamma$, has not been considered in the literature and the use of numerical methods, even if possible, is clearly beyond the scope of this paper. In order to obtain a first educated guess, we have followed the analytical solutions derived in the limit of small distortions in~\cite{Danese:1982} and the powerful constraints set by FIRAS on a possible chemical potential $\mu$ and a Comptonization $y$ distortion  which imply bounds up to $\Delta \rho_\gamma/\rho_\gamma\lesssim {\cal O}(10^{-4})$~\cite{Fixsen:1996nj}, depending on $T_{\rm res}$. 
These arguments lead to the constraint shown in Fig.~\ref{fig:HP} labelled ``CMB distortions'', which is the strongest requirement in the HP mass range neV$\,-\,2\times 10^{-4}$ eV. 

The lower limit of the CMB constraints, $m_{\gamma'}\sim {\rm neV}$, corresponds to resonances happening around the onset of recombination. Smaller HP masses suffer resonant transitions around this epoch because neutral Hydrogen contributes to $m_\gamma^2$ with a negative quantity which increases with time and eventually makes $m^2_\gamma=0$ by compensating the contribution of free electrons. 
HPs with masses down to $\sim 2\times 10^{-14}$ eV have their resonance around this epoch~\cite{Mirizzi:2009iz}.
It is very likely that CMB distortions will be produced for sub-neV masses, but since we cannot ascertain the fate of the low energy photons injected during recombination, we cannot use the results of~\cite{Danese:1982} and we must leave the study of this region for a future detailed numerical analysis.  
It is intriguing that a possible small distortion of the CMB spectrum due to HP CDM can in principle appear in the next generation of CMB spectral probes, such as PIXIE~\cite{Kogut:2011xw}. We believe, these signatures can be quite distinctive and strengthen the physics case for such probes, since they can help identifying the nature of the DM.   

There is however a simple way of constraining the kinetic mixing of CDM HPs in this low energy region. 
At late recombination, the density fluctuations are already imprinted in the CMB and they provide us with the estimate~\eqref{eq:observedDM} for the DM density. Since this value agrees roughly with the average DM density observed today we cannot allow the resonant depletion of HPs into photons at this epoch or any later, thus requiring  $\tau_2\lesssim 1$. This bound is depicted also in Fig.~\ref{fig:HP} and labelled as ``$\tau_2>1$''. 
In order to compute it, we have used the model for $m_\gamma^2$ provided in~\cite{Mirizzi:2009iz}.

Hidden photons with masses below twice the electron mass, can only decay via an electron loop into three photons. This decay rate is extremely suppressed for low mass HPs, $\Gamma_{3\gamma}\propto \alpha^4 m_{\gamma'}^9/m_e^8$~\cite{Redondo:2008ec}. 
However, for the most massive HPs considered here the decay can be effective. 
Imposing that the population of decay photons is smaller than the diffuse X-ray backgrounds one can constrain the cold HP population. This was done in~\cite{Redondo:2008ec} and we reproduce their bounds, labelled ``X-rays'', as the rightmost boundary of our HP CDM region in Fig.~\ref{fig:HP}.

\subsection{Thermal population of HPs}

During the resonance a thermal population of HPs can be generated without interfering with the arguments above. 
The role of this population as DM has been addressed in~\cite{Redondo:2008ec} (which excludes the region labelled ``Cosmology of thermal HP DM'' in Fig.~\ref{fig:HP}) and, as dark radiation, i.e.~contributing to the number of effective neutrinos, in~\cite{Jaeckel:2008fi}. 
This conversion also produces distortions of the CMB spectrum, which can be constrained by FIRAS. These arguments give rise to the bounds labelled ``CMB'' in Fig.~\ref{fig:HP}.

These thermal contributions are very much constrained by astrophysics and cosmology and on the verge of detectability by solar HP searches or 
{searches for} the contribution of the $\gamma'\to 3\gamma$ decay to the interstellar diffuse photon background. 

One could ask the question whether it is possible to have HP CDM and also explain the small excess of $N_{\nu}^{\rm eff}$ observed by WMAP7 and other CMB and large scale structure probes. 
The situation is a bit tricky since, for the parameters required to have the latter effect ($m_{\gamma'}\sim \mathcal{O}({\rm meV})$, $\chi\sim10^{-6}$)~\cite{Jaeckel:2008fi}, the CDM evaporation is huge, $\tau_2\gg 20$, and the required absorption of the CDM energy decreases $N_\nu^{\rm eff}$ opposing the first effect. Further studies are required to see if such a scenario is plausible, but it seems complicated to avoid the BBN bound coming from the dilution of the baryon density. 

{
\subsection{Indirect observational constraints on HPs}
For ALPs we have found powerful constraints from ALP decay and one may wonder if similar constraints exist for HPs.
Indeed, as mentioned above, HPs can decay via $\gamma'\to 3\gamma$ and this constrains the high mass region in Fig.~\ref{fig:HP}.

Similar to the ALP case one may also wonder about constraints from photon and cosmic ray propagation.
Photon propagation is essentially unaffected by a HP dark matter background since the combined photon-HP equations of motion are still linear
and consequently the superposition principle holds. In other words photons pass right through the HP background without interacting.

Cosmic rays on the other hand could scatter of the HPs via Compton scattering, $q+\gamma'\rightarrow q+\gamma$. However,
for the relevant values of the kinetic mixing parameter the cross section is too small to have a significant effect. 
}

\subsection{Direct experimental and observational constraints on HPs}

In Fig.~\ref{fig:HP} we have also displayed the existing experimental bounds on the existence of HPs which do not rely on HPs being DM.
The bounds labelled ``Solar lifetime'' and ``CAST'', coming from the non-observation of HP emission from the Sun, exclude a large portion of parameter space~\cite{Redondo:2008aa}. It is clear that improving the sensitivity of future searches of solar HPs one has access to new parameter space in which HPs can be CDM~\cite{Gninenko:2008pz}. The solar hidden photon search (SHIPS)~\cite{Schwarz:2011gu} in the Hamburg Observatory is currently exploring the sub-eV mass region greatly improving over the previous CAST experiment and will soon publish results.

Light-shining-through-walls experiments are also a powerful tool in the search for hidden photons~\cite{ITEP-48-1982}, the current best bound being provided by ALPS~\cite{Ehret:2010mh}, and shown in Fig.~\ref{fig:HP}. 
As in the case of ALPs they currently do not probe the DM region, but significant future improvements are expected~\cite{Redondo:2010dp}. Indeed, here the microwave regime is particularly promising~\cite{Jaeckel:2007ch,Caspers:2009cj}, with several new experiments already taking data~\cite{Povey:2010hs,Betz:2011}.

Finally, tests of Coulomb's law provide strong constraints in the low mass region, see Fig.~\ref{fig:HP}. 
Although these bounds do not reach into the HP DM region one can hope that improvements can be made since the best experiments~\cite{Williams:1971ms} on this are more than 40 years old.

\section{\label{sec:direct}Direct searches with haloscopes}

One well-known tool to search for axion dark matter are so-called axion haloscopes~\cite{Sikivie:1983ip}.
Let us briefly recap the basic principle of a haloscope. Using the abundance of axions all around us the task at hand is
to exploit their coupling to photons and to convert those axions 
into photons which can be detected. For axions this can be achieved by utilizing off-shell photons in the form of a strong magnetic field.
Moreover, this conversion can be made more efficient by employing a resonator, resonant at the frequency corresponding to the energy of
the produced photons.
The energy of the outgoing photons is equal to the energy of the incoming axions. As the DM particles are very cold their energy is dominated by their
mass. For axions in the natural dark matter window this mass is in the $1-100\,\mu{\rm eV}$ range corresponding to microwave photons.
A number of experiments of this type have already been done~\cite{DePanfilis:1987dk,Wuensch:1989sa,Hagmann:1990tj,Asztalos:2001tf,Asztalos:2009yp} and further improvements are 
underway~\cite{arXiv:1110.2180,Asztalos:2011bm,Heilman:2010zz}.

\subsection{Axion-like particles}
For axion-like particles the experiment proceeds exactly as in the axion case~\cite{Sikivie:1983ip} (and very similar to the hidden photon case described below).
The important point is, however, that now even bounds which do not reach the predictions of axion models (axion band in 
Fig.~\ref{fig:ALPDM}) become meaningful since they 
test viable models.

For completeness we recap. 
In the axion case, the power output of a cavity of volume $V$, quality $Q$ and coupling $\kappa$ to the detector is
\begin{equation}
P_{\rm out}=\kappa g^2 V |\mathbf{B}_{0}|^{2}\rho_{0}{\mathcal G}_{\rm axion}\frac{1}{m_{a}}Q,
\end{equation}
with $\rho_0$ the local axion CDM energy density, $|\mathbf{B}_{0}|$ the strength of the magnetic field and
\begin{equation}
{\mathcal{G}}_{\rm axion}=\frac{\left(\int dV\, \mathbf{E}_{\rm cav}\cdot \mathbf{B}_{0}\right)^2}{ |\mathbf{B}_{0}|^{2}V\int dV\, |\mathbf{E}_{\rm cav}|^{2}}.
\end{equation}
For cylindrical cavities in the TM$_{010}$ mode as, e.g., used in the ADMX experiment, ${\mathcal{G}}_{\rm axion}=0.68$.
The currently excluded region from various axion haloscopes is shown as gray area in Fig.~\ref{fig:ALPDM}.
It already excludes a part of the ALP CDM parameter space.

\subsection{Hidden photons}\label{hidcavity}

Microwave cavity experiments looking for relic axions could also be used to constrain and search for the hypothetical cold HP condensate that we have discussed above.
Starting from Eq.~\eqref{HPlagrangian} we can follow the usual route and trade the kinetic mixing term for a mass mixing by performing a shift
of the HP field $X\rightarrow X-\chi A$.
Neglecting terms of second order in $\chi$ the equation of motion for the photon field $A$ then reads,
\bb 
\label{precavity}
\partial_\mu \partial^\mu A^{\nu}=\chi m_{\gamma'}^2X^{\nu}.
\ee 
We can therefore see that the hidden photon field acts as a source for the ordinary photon.

Let us first determine the strength of this source.
As discussed above we can take $X^{0}=0$. For the spatial components we write ${\mathbf{X}}$. 
The energy density in the hidden photons is equal to the dark matter density. Therefore we have,
\begin{equation}
\rho_{\rm CDM}=\frac{m^{2}_{\gamma^{\prime}}}{2}|{\mathbf{X}}|^2.
\end{equation}

At this point a comment concerning the direction of the HP field is in order. 
In the discussion above it was assumed that the HP field is homogeneous in space and points in the same direction everywhere. 
Due to structure formation the DM is clumped. A common estimate for the DM energy density on Earth is therefore the 
energy density in the galactic
halo,
\begin{equation}
\rho_{\rm CDM, local}\sim \frac{0.3\,{\rm GeV}}{\rm cm^3}\gg\rho_{\rm CDM, average} \sim \frac{{\rm keV}}{\rm cm^{3}}.
\end{equation}
As we can see the local density is much higher than the average density, signifying that structure formation clearly is very important.
This raises the question whether structure formation also influences the direction of the HP condensate.
To answer this question one would have to study structure formation for a vector field like our HP field. 
This is beyond the scope
of the current paper. Instead we will consider two possible scenarios:
\begin{itemize}
\item[(a)] The direction of the HP field is (essentially) unaffected by structure formation and all 
HPs point in the same direction (at least for a suitably big region of space).
\item[(b)] The direction of HPs behave like a gas of particles with the vector pointing in random directions.
\end{itemize}

In scenario (a) the HP direction is characterized by a fixed vector $\hat{n}$, whereas in the case (b) we have to average
the final result over all directions for $\hat{n}$. 
With this understood, let us write,
\begin{equation}
\mathbf{X}(\mathbf{x})=\hat{\mathbf{n}} \frac{\sqrt{2\rho_{0}}}{{m_{\gamma'}}},
\end{equation} 
with $\rho_{0}$ the dark matter density on earth.

Let us now return to our cavity experiment.
The photon field $A$ inside the cavity can be expanded in terms of the cavity modes, 
\bb 
\mathbf{A}(\mathbf{x})
=\sum_i \alpha_i \mathbf{A}^{\rm cav}_i(\mathbf{x}), \ \ \ \ \  \int d^3\mathbf{x} |\mathbf{A}^{\rm cav}_i(\mathbf{x})|^2=C_i,
\ee 
with $C_i$ the normalisation coefficients.
Using this expansion and including losses in the cavity we obtain for the expansion coefficients,
\begin{equation}
\left(\frac{d^{2}}{dt}+\frac{\omega_{0}}{Q}\frac{d}{dt}+\omega^{2}_{0}\right) \alpha_{i}(t)=b_{i}\exp(-i\omega t),
\end{equation}
with $\omega_0$ the frequency of the cavity and $Q$ its quality factor.
The driving force $b_{i}$ can be written as
\bb 
b_{i}=\frac{\chi m^{2}_{\gamma^{\prime}}}{C_{i}}  
\int dV {\mathbf{A}}^{\star}_{i}(\mathbf{x})\cdot\mathbf{X}(\mathbf{x})
\ee 
and the frequency is given by the energy of the HPs,
\begin{equation}
\omega=E_{\gamma^{\prime}}\approx m_{\gamma^{\prime}}.
\end{equation}

The asymptotic solution for the cavity coefficients is then,
\begin{equation}
\alpha_{i}(t)=\alpha_{i,0}\exp(-i\omega t)=\frac{b_{i}}{\omega^2_{0}-\omega^{2}-i\frac{\omega\omega_{0}}{Q}}\exp(-i\omega t).
\end{equation}

Finally, the power emission of the cavity is related to the energy stored and the quality factor of the cavity as
\bb 
\mathcal P_{\rm out}=\kappa\frac{U}{Q}\omega_0,\ee 
where $\kappa$ is the coupling of the cavity to the detector and
\bb 
\label{energy}
U=\frac{|\alpha_{i,0}|^2 \omega^2}2\int d^3\mathbf{x}\, |\mathbf{A}^{\rm cav}_{i}(\mathbf{x})|^2. 
\ee 
Replacing in this equation and evaluating at resonance
$\omega_0= m_{\gamma'}$ \footnote{As in the axion case using this condition requires that the $Q$ is not too
large since after structure formation the DM particles move with different velocities of order $300$~km/s$\sim 10^{-3} c$.
This restricts the maximal usable $Q$ to be $\lesssim 10^6$. If axions form a Bose-Einstein condensate (see~\cite{Sikivie:2009qn,Erken:2011dz} and Sect.~\ref{essentials}) or the galactic exhibits special structures as suggested in~\cite{Sikivie:1992bk,Duffy:2008dk}, the velocity distribution could have significantly narrower structures potentially allowing to utilize a cavity with higher~$Q$. Moreover, in this case one could also benefit from using narrow bandwidth techniques. Strong magnetic fields typically limit the maximal $Q\lesssim 10^{6}$. Therefore, the former only works for hidden photons, where no magnetic field is needed and one can use superconducting cavities. The latter however, would work for both HPs and ALPs and has indeed been employed in ADMX~\cite{Hoskins:2011iv}.}
we find
\bb 
\mathcal P_{\rm out}=\kappa \chi^2 m_{\gamma'} \rho\, Q\,  V\, \mathcal G,
\ee 
where the geometric factor 
$\mathcal G$ is defined as
\bb 
\mathcal G
= \frac{\left|\int dV{\mathbf{A}}^{*\rm cav}(\mathbf{x})\cdot \hat{\mathbf{n}}\right|^2}{V\, \int d^3\, \mathbf{x} |\mathbf{A}^{\rm cav}(\mathbf{x})|^2}.
\ee

In a cavity ${\mathbf{E}}=\omega {\mathbf{A}}$. This geometry factor has exactly the same form as in the axion case (cf.~\cite{Asztalos:2001tf} and the subsection above)
but with the direction $\hat{\mathbf{n}}$ replacing the direction of the external magnetic field $\bf{B}_{0}$ in the axion case.
Accordingly,
\begin{equation}
\mathcal{G}=\mathcal{G}_{\rm axion}\cos^{2}(\theta),
\end{equation}
where $\theta$ is the angle between the magnetic field direction used in the axion case (this is usually chosen to be optimal) and the
direction $\hat{\mathbf{n}}$ of the hidden photon field. 

We can now use this formula to constrain the HP CDM with the present microwave cavity searches for 
axions~\cite{DePanfilis:1987dk,Asztalos:2001tf,Hagmann:1990tj,Wuensch:1989sa,Asztalos:2009yp}. 

For scenario (a) the result depends on the relative orientation of the cavity at the time of the measurement with respect to an a priori unknown direction of the HP field. This requires detailed knowledge of the timing of the experiment etc.
A conservative estimate for the sensitivity can be obtained, however, by assuming that all directions in space are equally likely
and taking a value for $\cos^2(\theta)$ such that the real value is bigger with $95\%$ probability. The corresponding value {is $\cos^{2}(\theta)=0.0025$.}
For situation (b) one can average over all possible directions and {obtain $\langle \cos^{2}(\theta)\rangle=1/3$.}

The results of the analysis are shown in Fig.~\ref{fig:HP}, under the label ``Haloscope Searches''. Blue shows scenario (a) with our
simplified estimate and red scenario (b). With a more careful analysis of the experiment, including the timing information, the sensitivity
in scenario (a) would be very similar to the one in scenario (b).

\subsection{Non-resonant searches}
Both for ALP DM and for HP DM the coupling to photons can be orders of magnitude bigger than in the axion case.
At the same time it is desirable to explore a bigger mass range. For a first broadband search one may therefore be prepared to give up the resonant enhancement in the cavity.  
Indeed the formulas in Sect.~\eqref{hidcavity} up to Eq.~\eqref{energy} are valid also off resonance. One can obtain the general result 
by replacing\footnote{Note $Q$ may be a function of $\omega$.}  
\begin{equation}
Q\rightarrow \frac{Q(\omega_{0})}{Q^{2}(\omega)}\frac{\kappa(\omega)}{\kappa(\omega_{0})}\frac{\omega_{0}}{\omega}
\left|\frac{\omega^2}{\omega^2_{0}-\omega^{2}-i\frac{\omega\omega_{0}}{Q(\omega)}}\right|^2.
\end{equation}
In principle one can therefore explore a wide range off masses with a single fixed frequency cavity (at the price of some loss of sensitivity).
However, it should be noted that in order to do so one also has to have a receiver sensitive in a sufficiently wide range of frequencies and also check for signals off the resonance of the cavity.

\section{\label{conclusions}Conclusions}

Vacuum misalignment in the very early universe is a very generic mechanism capable of producing 
a cold dark matter condensate from any very light, but massive (quasi-)stable bosonic field. Clearly, its self-interactions and
its interactions with the SM should be very weak if it is to survive and to be a candidate for the observed cold dark 
matter today -- in other words their particle excitations should be very weakly interacting slim particles (WISPs). 
 
We found that for the most prominent and theoretically well-motivated WISP candidates -- axion-like particles (ALPs) 
and hidden photons (HPs),    
whose dominant interaction with the standard model arise from couplings to photons -- a huge region in parameter space 
spanned by their masses and their photon coupling can give rise to the observed cold dark matter. 

WISPy CDM coupled to photons is particularly interesting because it can be probed by both astrophysical observations as well as laboratory experiments.
Figs.~\ref{fig:ALPDM} and \ref{fig:HP} show that sizable regions of this parameter space have already been excluded
by experiments searching for WISPs without relying on them being dark matter. 
In particular, helioscopes -- experiments looking for WISPs produced in the sun -- have already probed regions that allow WISPy DM and will
improve further in the near future.
Soon also purely laboratory-based light-shining-through-walls experiments (in the optical as well as in the microwave regime) will reach a sensitivity that will allow them to test hidden photon DM with masses in the $\mu {\rm eV} -{\rm meV}$ region.  
 
The possibility that WISPs can form all or part of DM allows for additional search strategies. One way to exploit this
is via haloscopes which already test both ALPs and hidden photons extremely sensitively in the $\mu {\rm eV}$ mass region.
Moreover, searches for light emitted from decaying ALPs already provide interesting constraints in a large range of masses from meV to MeV
and will hopefully further improve in the future.

Despite these bright prospects for future searches it is clear that huge areas of parameter space allowing for WISPy CDM are so far unexplored and demand not only new searches but also new search strategies -- a challenge for both experiment and phenomenology.  

\section*{\label{Acknowledgements}Acknowledgements}

P.~A. was supported by the Humboldt Foundation, M.~G. was supported by the SFB 676 and by the ERC advanced grant 226371. D.~C. and J.~R.  acknowledge partial support from the European Union FP7 ITN INVISIBLES (Marie Curie Actions, PITN-GA-2011-289442). We would like to thank Christian Byrnes, Michele Cicoli, Dieter Horns, Axel Lindner, and Gray Rybka for interesting discussions. Moreover, the authors are very grateful to Valery Rubakov and Christof Wetterich for pointing out an error in the cosmological evolution of a vector field.

%

\end{document}